\documentclass[utf8,bibauthoryear]{aa}
\usepackage{graphicx}
\usepackage{txfonts}
\usepackage{siunitx}
\usepackage[utf8]{inputenc}
\usepackage{natbib}
\usepackage{caption}
\usepackage{subcaption}
\usepackage{amsmath}
\usepackage{array} \usepackage{bm} \usepackage[pdftex, hidelinks]{hyperref}
\begin{document}

\title{Non-ideal self-gravity  and cosmology: the importance  of correlations in
  the dynamics of the large-scale structures of the Universe}

\author{P.  Tremblin \inst{1}\thanks{\email{pascal.tremblin@cea.fr}}
  \and G. Chabrier \inst{2,3}
  \and T. Padioleau \inst{1}
  \and S. Daley-Yates \inst{1}
}

\institute{Maison  de   la  Simulation,   CEA,  CNRS,  Univ.   Paris-Sud,  UVSQ,
  Universit\'e Paris-Saclay, F-91191 Gif-sur-Yvette, France\
  \and Ecole Normale Sup\'erieure de Lyon, CRAL, UMR CNRS 5574, 69364 Lyon Cedex
  07, France\
  \and Astrophysics Group, University of Exeter, EX4 4QL Exeter, UK }

\date{Received \#\#\# \#\#\, 2019; accepted \#\#\# \#\#, 2019}

% Abstract of the paper
\abstract  {}  {Inspired  by  the   statistical  mechanics  of  an  ensemble  of
  interacting particles (BBGKY hierarchy), we propose to account for small-scale
  inhomogeneities  in  self-gravitating  astrophysical   fluids  by  deriving  a
  non-ideal  Virial  theorem  and   non-ideal  Navier-Stokes  equations.   These
  equations  involve  the pair  radial  distribution  function (similar  to  the
  two-point correlation function used to characterize the large-scale structures
  of the Universe), similarly to the interaction energy and equation of state in
  liquids.  Within this framework, small-scale  correlations lead to a non-ideal
  amplification of the gravitational interaction energy, whose omission leads to
  a missing mass problem, e.g., in galaxies and galaxy clusters.}
{We propose to  use a decomposition of the gravitational  potential into a near-
  and far-field  component in order to  account for the gravitational  force and
  correlations  in the  thermodynamics properties  of the  fluid.  Based  on the
  non-ideal  Virial theorem,  we  also  propose an  extension  of the  Friedmann
  equations in the  non-ideal regime and use numerical  simulations to constrain
  the contribution  of these correlations  to the expansion and  acceleration of
  the Universe.}
{We estimate the non-ideal amplification factor of the gravitational interaction
  energy of  the baryons  to lie  between 5 and  20, potentially  explaining the
  observed value of the Hubble  parameter (since the uncorrelated energy account
  for $\sim$  5\%).  Within  this framework, the  acceleration of  the expansion
  emerges naturally because  of the increasing number  of sub-structures induced
  by gravitational  collapse, which  increases their  contribution to  the total
  gravitational  energy.  A  simple estimate  predicts a  non-ideal deceleration
  parameter $q_{\rm ni}\simeq$  -1; this is potentially  the first determination
  of  the observed  value based  on an  intuitively physical  argument. We  also
  suggest that small-scale gravitational interactions in bound structures (spiral
  arms or  local clustering) could yield  a transition to a  viscous regime that
  can lead  to flat  rotation curves.   This transition  could also  explain the
  dichotomy between (Keplerian) LSB elliptical galaxy and (non-Keplerian) spiral
  galaxy  rotation profiles.   Overall, our  results demonstrate  that non-ideal
  effects induced  by inhomogeneities  must be  taken into  account, potentially
  with our formalism, in order  to properly determine the gravitational dynamics
  of galaxies and the larger scale universe.}  {}


\titlerunning{Non-ideal self-gravity}
\authorrunning{P. Tremblin}

\maketitle

%
%-------------------------------------------------------------------

   \section{Introduction} \label{sec:introduction}

   Astrophysical  flows are  by nature  multi-scale systems  exhibiting a  large
   range   of    dynamical   regimes    whose   understanding    remains   quite
   challenging. Since  the 70s/80s, hydro  or N-body high  performance numerical
   simulations  have  been key  tools  to  understand self-gravitating  systems.
   However, a  proper understanding  of collective  effects in  such simulations
   remain  challenging because,  notably,  of  numerical artifacts.   Therefore,
   theory  remains vital  for assessing  the genuine  validity of  the numerical
   simulations.

   The theoretical description of an ensemble of particles is a challenging task
   that has motivated the development of  a major domain in physics: statistical
   mechanics. The statistical  descriptions of a gas and a  solid are relatively
   easy in  two limit cases:  perfect disorder,  often referred to as molecular
   chaos for ideal  gases, or perfect order for ordered  solids. The description
   of liquids, or  ill-condensed systems, however, is much more  complex and has
   only  been possible  after  the  pioneering work  of  Bogolyubov, Born,  Green,
   Kirkwood,    and    Yvon,    referred    to   as    the    BBGKY    hierarchy
   \citep[e.g.][]{yvon:1935,born:1946,bogoliubov:1946,kirkwood:1946}.     Within
   this framework,  the ensemble  of particles is  described by  the one-particle
   probability density function, the  two-particle probability density function,
   and so  on. Truncated  at the second  order, the system  is described  by the
   one-particle density  function and  the pair correlation  function, generally
   referred  to  as  the  radial distribution  function  in  liquid  statistical
   mechanics.   In the  case  of perfect  chaos  for an  ideal  gas, the  radial
   distribution function is  uniformly equal to 1, which means  that there is no
   correlation in  the fluid: the  probability of  finding two particles  at two
   given points  is equal  to the  product of the  probabilities of  finding one
   particle at these given points.

   Liquids differ  from an ideal gas  because of the presence  of inter-particle
   forces that  lead to  correlations at  short distances  and they  differ from
   solids because  of the lack  of (periodic)  long-range order (i.e.   the pair
   correlation function  tends to 1 at  long distance). For water,  the dominant
   effects responsible for these inter-molecular forces are dipolar interactions
   and  hydrogen bonding.   Many other  processes can  be at  play and  make the
   picture more complex. However, from a statistical point of view the knowledge
   of the radial distribution function is sufficient to derive the thermodynamic
   properties    of   the    fluid,   such    as   the    equation   of    state
   \citep[see][]{hansen:2006,aslangul:2006}.

   Most of the  developments of liquid statistical mechanics deeply  rely on the
   fact  that  the  interactions  at  play  are  short-range:  even  though  the
   interaction energy  is proportional to  the number  of particle pairs  in the
   fluid, only  the close  neighbors actually  matter.  This  allows to  find an
   extensive definition of  the interaction energy that does not  diverge in the
   thermodynamic limit. A noticeable exception  is the case of Coulombic fluids,
   which interact throughout the Coulomb  potential ($\propto 1/r$); large-scale
   convergence in that case is  insured by the neutralizing electron background.
   Unfortunately, this  (screening) property  is not  met by  a self-gravitating
   fluid since the gravitational force  is long-range and always attractive: the
   interaction  energy is  a priori  non-extensive and  requires possibly  a new
   extension  of statistical  mechanics  to  include non-extensive  hierarchical
   systems   \citep[see,  e.g.][]{devega:2002,pfenniger:2006}.    This  prevents
   astrophysicists from benefiting  from the insight found  from developments of
   statistical mechanics in  the presence of interactions.  For  this reason, at
   large scales such as galaxies or  the large-scale structures of the Universe,
   the gravitational  interactions of  unresolved structures in  the simulations
   are usually  ignored or  neglected, resulting  from the use  of an  ideal gas
   approximation   when  calculating   the  thermodynamic   properties  of   the
   fluid. Such  an approximation,  however, can lead  to inconsistencies  in our
   description of the Universe.
   %e.g.    the  missing   mass  problem   in  galaxies   and  galaxy   clusters
   %\citep{zwicky:1937,rubin:1980}  and the  need to  introduce ad-hoc  mass and
   %energy  in  the Friedmann  equations  in  order  to reproduce  the  observed
   %expansion and acceleration of the Universe.

   The present paper  aims at proposing to properly  apply statistical mechanics
   to gravitational  systems by  solving the  problem of  non-extensivity.  This
   leads  to  the  derivation  of  a  non-ideal  Virial  theorem  and  non-ideal
   Navier-Stokes equations to be  used for self-gravitating hydrodynamic fluids
   when,   as  is   the  case   within   the  universe,   correlations  due   to
   self-gravitating sub-structures are  present.  In Sect.~\ref{sec:liquid-eos},
   we recall  classical results of  statistical mechanics  in order to  define a
   non-ideal  equation of  state  (EOS)  in the  presence  of interactions.   In
   Sect.~\ref{sec:grav-eos},  we propose  a  way  to adapt  these  tools to  the
   long-range  gravitational   force  by   using  a  near-field   and  far-field
   decomposition  of  the potential.   In  Sect.~\ref{sec:appli},  we present  a
   semi-analytical example using polytropic stellar structures to illustrate the
   importance  of inhomogeneities  and correlations  when computing  the correct
   interaction energy in  the Virial theorem.  Then, we explore  the role of the
   gravitational  force and  the correlations  at  small scales  in galaxies  to
   explain  the observed  flat  rotation  curves.  Finally,  using  for now  the
   Newtonian  limit, we  define non-ideal  Friedmann equations  and explore  the
   possible  consequences of  these  non-ideal  effects in  the  context of  the
   expanding  and  accelerating  Universe.   In  Sect.~\ref{sec:conclusions}  we
   summarize our  conclusions, we  discuss the  limitations of  our work  and we
   propose ways to go further in exploring this potentially interesting idea.

   \section{Statistical mechanics of an ensemble of interacting particles}
   \label{sec:liquid-eos}

   \subsection{BBGKY hierarchy}\label{sec:bbgky}

   Following the work  of Bogolyubov, Born, Green, Kirkwood, and  Yvon, known as
   the BBGKY hierarchy, we introduce some notations of statistical mechanics. We
   consider  $N$ particles  of  mass $m$  within  a volume  $V$.  We can  define
   $P_s(\vec{r}_1,\vec{r}_2,...,  \vec{r}_s)$   the  probability  to   find  $s$
   particles at the  positions ($\vec{r}_1$,$\vec{r}_2$,..., $\vec{r}_s$). These
   probability density functions are normalized  by the total number of N-uplets
   $N^s$. Two  probability functions  play an  important role  when there  is no
   interaction   involving  more   than  two   particles:  $P_1(\vec{r})$,   the
   probability     to     find     one     particle     at     $\vec{r}$     and
   $P_2(\vec{r},\vec{r^\prime})$,  the  probability  to  find  one  particle  at
   $\vec{r}$  and  one  particle  at  $\vec{r^\prime}$.  These  two  probability
   functions can  be used to define  exactly fluid quantities involving  one and
   two particles from their N-body counterparts.

   For  instance, using  $P_1(\vec{r})$, we  can define  equivalently the  total
   acceleration $ \langle  \vec{a} \rangle$ of the fluid volume  $V$ and the sum
   of all the N-body particle accelerations:

   \begin{eqnarray}
     \langle   \vec{a}  \rangle   &=&  \sum_i   \vec{a}_i  \cr   &=&  N   \int_V
     \vec{a}(\vec{r}) P_1(\vec{r}) dV
   \end{eqnarray}

   The use of $P_1(\vec{r})$, here, involves to counting the number of particles
   in the volume $V$ since by definition the density inside the fluid volume is:

   \begin{equation}\label{eq:density}
     \rho(\vec{r}) = N P_1(\vec{r}).
   \end{equation}
   In the  case of a homogeneous  and isotropic fluid, $P_1(\vec{r})$  is simply
   equal to  $1/V$ for all  $\vec{r}$ and $\rho$ is  equal to $N/V$,  the number
   density.

   Similarly, one can define the interaction energy of the ensemble of particles
   $\langle  H_\mathrm{int} \rangle$:  by summing  all the  interaction energies
   between  distinct  pairs  of  particles  in  the  N-body  description  or  by
   integrating over $P_2(\vec{r},\vec{r^\prime})$ for the fluid volume.

   \begin{eqnarray}\label{eq:hint_nbody_fluid}
     \langle H_\mathrm{int}  \rangle &=&  \sum_{i<j} \phi(|\vec{r}_i-\vec{r}_j|)
     \cr  &=& \frac{N(N-1)}{2}  \iint_{V,V} \phi(|\vec{r}-\vec{r}^\prime|)
     P_2(\vec{r},\vec{r^\prime}) dV dV^\prime,
   \end{eqnarray}
   where $N(N-1)/2$  is the  number of  independent pairs in  the fluid.  In the
   homogeneous and isotropic case, $P_2(\vec{r},\vec{r^\prime})$ only depends on
   the distance modulus $|| \vec{r}-\vec{r^\prime}||$.  It is then convenient to
   define the radial distribution function

   \begin{equation}
     g(||\vec{r}-\vec{r^\prime}||) = V^2 P_2(\vec{r},\vec{r^\prime}),
      \label{eq:gr}
   \end{equation}
   or equivalently the  correlation function, $\xi(||\vec{r}-\vec{r^\prime}||)$,
   for a homogeneous isotropic fluid:

   \begin{eqnarray}
     \xi(||\vec{r}-\vec{r^\prime}||)   &=&   V^2(P_2(\vec{r},\vec{r^\prime})   -
     P_1(\vec{r})P_1(\vec{r^\prime}))\cr  &=& V^2  P_2(\vec{r},\vec{r^\prime}) -
     1\cr &=& g(||\vec{r}-\vec{r^\prime}||) -1.
     \label{eq:corr}
   \end{eqnarray}

   For  an   isotropic  fluid,   the  correlation   function  $\xi(r)$   is  the
   angle-averaged       value        of       $\xi(\vec{r})=\langle       \delta
   (\vec{r}^\prime)\delta(\vec{r}+\vec{r}^\prime)\rangle$,                  with
   $\delta(\vec{r})$  the  density  perturbation  defined  as  $\delta(\vec{r})=
   (\rho(\vec{r})-\langle      \rho     \rangle)/\langle      \rho     \rangle)$
   \citep[see][]{springel:2017}. Here, $\langle \cdot  \rangle$ denotes a volume
   average. The power  spectrum can then be defined as  the Fourier transform of
   $\xi(\vec{r})$:
   \begin{equation}
     \xi(\vec{r})    =    \frac{V}{(2\pi)^3}\int   |\delta_{\vec{k}}|^2    e^{-i
       \vec{k}\cdot\vec{r}}d^3k,
   \end{equation}
   which  is  commonly   used  in  cosmological  studies   to  characterize  the
   large-scale structures of the Universe \citep{peebles:1980,peacock:1999}.  In
   this  paper, we  will show  that  the correlation  function is  not a  simple
   diagnostic of  structures but that  it plays  an active role  determining the
   dynamics of the fluid.  For the sake  of similarity with the usual tools used
   in statistical physics,  we will use the radial  distribution function rather
   than the correlation function in our calculations.

\begin{figure}
   \begin{centering}
     \includegraphics[width=\linewidth]{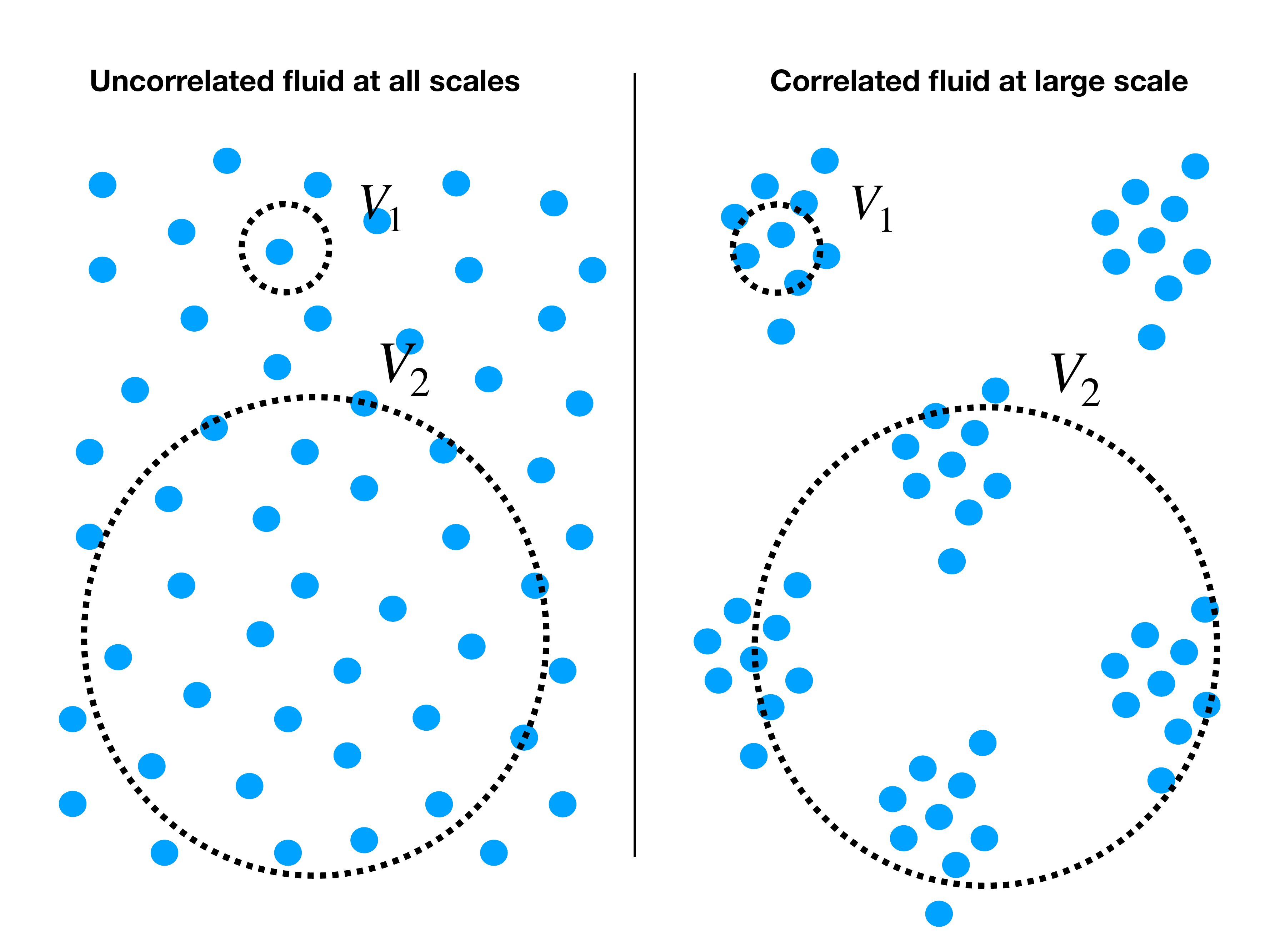}
     \caption{Schematic illustration of situations  when the ideal hypothesis is
       valid and  when it breaks  down. Left: the  fluid is uncorrelated  at all
       scales. Right: on the scale of the control volume $V_1$, the fluid can be
       considered locally as  inhomogeneous and uncorrelated while  on the scale
       of the  control volume  $V_2$, the  fluid can  be considered  as globally
       homogeneous and correlated.} \label{fig:distribution}
   \end{centering}
   \end{figure}

   It  is  easy  to see  that  in  the  absence  of correlation  in  the  fluid:
   $P_2(\vec{r},\vec{r^\prime})    =   P_1(\vec{r})P_1(\vec{r^\prime})$,    i.e.
   $\xi(||\vec{r}-\vec{r^\prime}||)=0$ and $g(||\vec{r}-\vec{r^\prime}||)=1$ for
   all $\vec{r}$ and $\vec{r^\prime}$ in the  case of a homogeneous fluid.  This
   situation is the limit case of an ideal fluid, often referred to as ``perfect
   molecular chaos'', for  which the radial distribution function is  equal to 1
   everywhere.   A schematic  representation of  such an  ideal and  homogeneous
   fluid is portrayed  in the left panel of  Fig.  \ref{fig:distribution}, where
   the blue dots represent point-like interactionless particles. In this example
   the fluid is homogeneous at all scales: the density inside a small ($V_1$) or
   large ($V_2$)  control volume is independent  of the location of  the control
   volume.

   In the  presence of interactions, the  density distribution may no  longer be
   homogeneous at all scales, e.g. in the presence of clustering in the particle
   distribution.   Such  a   situation  is   shown   in  the   right  panel   of
   Fig. \ref{fig:distribution}:

   \begin{itemize}
   \item for a small control volume  $V_1$, the distribution of particles within
     the    volume     is    ideal     and    there    is     no    correlation:
     $P_2(\vec{r},\vec{r^\prime})  =  P_1(\vec{r})P_1(\vec{r^\prime})$.  However,
     the density will depend  on the location of the control  volume, e.g. if it
     lies  within a  cluster or  outside.  This means  that inhomogeneities  are
     resolved and the fluid is ideal (i.e. uncorrelated) but inhomogeneous.
   \item for a large control volume  $V_2$, the distribution of particles within
     the  volume  is  not  ideal  and there  are  correlations  induced  by  the
     clustering   inside    the   volume:    $P_2(\vec{r},\vec{r^\prime})   \neq
     P_1(\vec{r})P_1(\vec{r^\prime})$. However  the density  will not  depend on
     the location  of the control  volume provided  that the control  volume is
     sufficiently large. In  that case, the inhomogeneities are  not resolved and
     the fluid is non-ideal (correlated) and homogeneous.
   \end{itemize}

   This   distinction  is   very  important   for  the   correct  treatment   of
   inhomogeneities (correlations)  in a  fluid. Since it  depends on  the scale,
   this is of prime important for astrophysical flows, that can only be observed
   or simulated at large scales.  The main consequence is that one cannot simply
   average  over large  volumes  as this masks the  complexity  of the  density
   inhomogeneities:  these inhomogeneities  then result  in correlations  in the
   fluid that  requires a change  to the dynamical  equations. This can  be seen
   with the computation  of the interaction energy,  which depends intrinsically
   on the distance between particles (see Eq.~\ref{eq:hint_nbody_fluid}). Within
   a  statistical  (fluid)  description, the  interaction  energy  fundamentally
   involves  the  pair   correlation  function,  $P_2(\vec{r},\vec{r^\prime})\ne
   P_1(\vec{r})P_1(\vec{r^\prime})$.  In  a simulation assuming an  ideal fluid,
   the  correlations  will  be  missed,  i.e.  the  fluid  at  small  scale  may
   erroneously be  considered as ideal  if correlations are present, yielding an
   erroneous estimate  of the  interaction energy.   The proper  calculation of
   this  latter must  thus  be  done accurately,  making  sure  the small-scale
   interaction   energy  is   properly   accounted   for,  potentially   through
   Eq.~\ref{eq:hint_nbody_fluid} and the use of a non-ideal framework.

   \subsection{Virial theorem for correlated fluids}\label{sec:virial}

   For a  system of interacting  particles of mass $m$,  the N-body form  of the
   Virial theorem can be stated as:

   \begin{equation}
     -\sum_i   m   \vec{v}_i^2   +\frac{1}{2}\sum_i   \frac{d^2   I_i}{dt^2}   =
     \sum_{i,j,i\neq j} \vec{F}_{j\rightarrow i} \cdot \vec{r}_i
   \end{equation}
   where $I_i = m r_i^2$ is the moment  of inertia of the particle i. In a fluid
   (statistical) approach, the equivalent form reads:

   \begin{multline}
     - N  \int_V  m  \vec{v}^2(\vec{r})  P_1(\vec{r})  dV+\frac{1}{2}  N  \int_V
     \frac{d^2   I}{dt^2}(\vec{r})   P_1(\vec{r})dV   \\   =   N(N-1)\iint_{V,V}
     \vec{F}_{\vec{r}^\prime     \rightarrow      \vec{r}}     \cdot     \vec{r}
     P_2(\vec{r},\vec{r}^\prime)dVdV^\prime.
   \end{multline}
   Assuming    a   force    deriving    from   a    potential   $\phi    \propto
   |\vec{r}-\vec{r}^\prime|^\alpha$,  the last  integral can  be symmetrized  by
   using $P_2(\vec{r},\vec{r}^\prime)=P_2(\vec{r}^\prime,\vec{r})$ to get
   \begin{eqnarray}
     &  N(N-1)& \iint_{V,V}  \vec{F}_{\vec{r}^\prime \rightarrow  \vec{r}} \cdot
     \vec{r}  P_2(\vec{r},\vec{r}^\prime)dVdV^\prime  \cr  &=&  \frac{N(N-1)}{2}
     \iint_{V,V}    \vec{F}_{\vec{r}^\prime     \rightarrow    \vec{r}}    \cdot
     (\vec{r}-\vec{r}^\prime) P_2(\vec{r},\vec{r}^\prime) dVdV^\prime  \cr &=& -
     \alpha    \frac{N(N-1)}{2}    \iint_{V,V}    \phi(|\vec{r}-\vec{r}^\prime|)
     P_2(\vec{r},\vec{r}^\prime)dVdV^\prime
   \end{eqnarray}
   in    which    we   recognize    the    interaction    energy   defined    in
   Eq.~\ref{eq:hint_nbody_fluid}.  The Virial theorem takes now the classic form
   \begin{equation}
     \langle   m  v^2   \rangle   -\alpha  \langle   H_\mathrm{int}  \rangle   =
     \frac{\langle d^2I/dt^2 \rangle}{2}
   \end{equation}
   in  which we  note  the presence  of  the pair  correlation  function in  the
   interaction energy.

   Similarly, we  can define  the total  mechanical energy  $E_m$ of  the N-body
   ensemble of particles
   \begin{eqnarray}
     E_m &=& \langle  m \vec{v}^2/2 \rangle +  \langle H_\mathrm{int} \rangle\cr
     &=&     \frac{1}{2}      \sum_i     m     \vec{v}_i^2      +     \sum_{i<j}
     \phi(|\vec{r}_i-\vec{r}_j|),
   \end{eqnarray}
   whose fluid counterpart is given by
   \begin{equation}
     E_m   =    \frac{N}{2}   \int_V   m    \vec{v}^2(\vec{r})P_1(\vec{r})dV   +
     \frac{N(N-1)}{2}                 \iint_{V,V}\phi(|\vec{r}                 -
     \vec{r}^\prime|)P_2(\vec{r},\vec{r}^\prime) dVdV^\prime.
   \end{equation}
   If   we    neglect   correlations    in   the   interaction    energy,   then
   $P_2(\vec{r},\vec{r^\prime})   =    P_1(\vec{r})P_1(\vec{r^\prime})$,   which
   yields, in the limit $N\gg 1$
   \begin{eqnarray}\label{eq:hint_ideal}
     \langle   H_\mathrm{int}   \rangle  &\approx&   \frac{N^2}{2}   \iint_{V,V}
     \phi(|\vec{r}-\vec{r}^\prime|)
     P_1(\vec{r})P_1(\vec{r}^\prime)dVdV^\prime\cr                     &\approx&
     \frac{1}{2}\iint_{V,V}                       \phi(|\vec{r}-\vec{r}^\prime|)
     \rho(\vec{r})\rho(\vec{r}^\prime)dVdV^\prime\cr                   &\approx&
     \frac{1}{2}\int_{V} \rho(\vec{r})\Phi(\vec{r})dV,
   \end{eqnarray}
   where  we  have  defined  the  mean-field  potential  $\Phi(\vec{r})=\int_{V}
   \phi(|\vec{r}-\vec{r}^\prime|)  \rho(\vec{r}^\prime)dV^\prime$, which  can be
   computed with the Poisson equation for the gravitational or the electrostatic
   force, $\nabla^2  \Phi(\vec{r}) \propto  \rho(\vec{r})$.  However,  we stress
   that the possibility to define this  mean-field potential is only possible in
   the  absence of  correlations: as  soon as  $P_2(\vec{r},\vec{r^\prime}) \neq
   P_1(\vec{r})P_1(\vec{r^\prime})$,  the use  of a  mean-field and  the Poisson
   equation is  not correct as  it will ignore  the correlations present  in the
   system.

   \begin{figure}
   \begin{centering}
     \includegraphics[width=\linewidth]{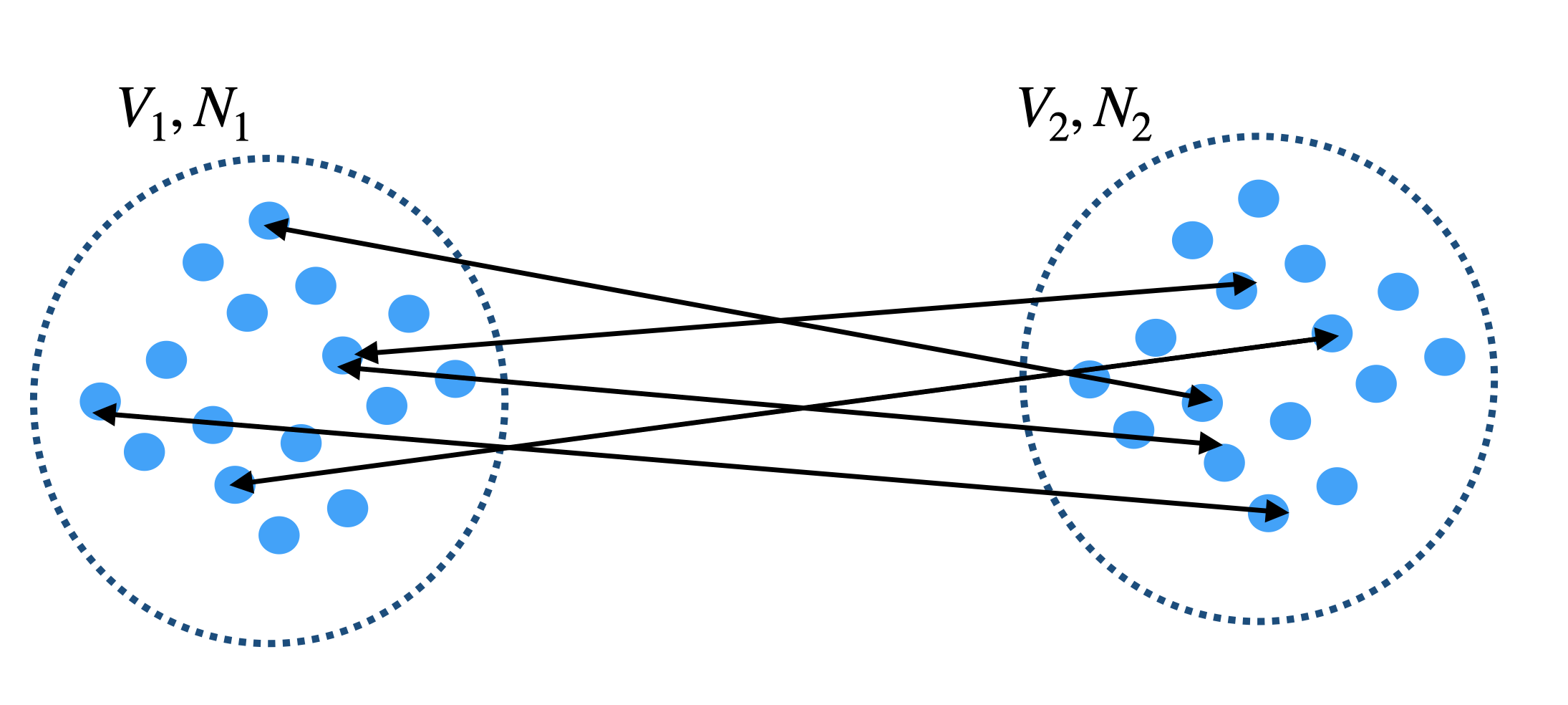}
     \caption{Schematic illustrating the interactions  between two fluid volumes
       used  for the  derivation of  Newton's laws  for pressureless  correlated
       fluids.} \label{fig:newton_fluid}
   \end{centering}
   \end{figure}

   \subsection{Newton's laws of motion for pressureless correlated fluids}
   \label{sec:newton}

   At a  more fundamental level,  we can also  show that correlations  should be
   taken into account in Newton's laws for a (pressureless) correlated fluid. We
   consider now an ensemble  of $N_1$ particles of mass $m_1$  in a volume $V_1$
   interacting  with $N_2$  particles  of  mass $m_2$  in  a  volume $V_2$  (see
   Fig.~\ref{fig:newton_fluid}). In  an N-body approach,  the second law  can be
   written
   \begin{equation}
       \sum_i m_1 \vec{a}_i = \sum_{j,i} \vec{F}_{j\rightarrow i},
   \end{equation}
   with $i$  in $[0,N_1]$ and  $j$ in $[0,N_2]$. The  Newton's second law  for a
   pressureless correlated fluid is then given by
   \begin{equation}
       N_1\int_{V_1}  m_1 \vec{a}(\vec{r})P_1(\vec{r})dV_1  = N_1  N_2\iint_{V_1,
         V_2}  \vec{F}_{\vec{r}_2\rightarrow  \vec{r}_1}P_2(\vec{r}_1,\vec{r}_2)
       dV_1 dV_2,
   \end{equation}
   which  we write  in a  more compact  way as  $\langle m_1\vec{a}  \rangle_1 =
   \langle \vec{F}  \rangle_{2\rightarrow1}$. Note that we  have generalized the
   two-point probability density  function to different volumes  $V_1$ and $V_2$
   and it is normalized by $N_1 N_2$.

   The  first  law for  an  isolated  fluid of  volume  $V_1$  is then  $\langle
   m_1\vec{a} \rangle_1 =  \langle \vec{F} \rangle_{1\rightarrow1} =  0$ and can
   be   deduced  from   the   second   law  by   taking   $V_2=V_1$  and   using
   $P_2(\vec{r}_1,\vec{r}_1^\prime)   =   P_2(\vec{r}_1^\prime,\vec{r}_1)$   and
   $\vec{F}_{\vec{r}_1\rightarrow               \vec{r}_1^\prime}              =
   -\vec{F}_{\vec{r}_1^\prime\rightarrow  \vec{r}_1}$.  Similarly,  one can  get
   the third  law $\langle \vec{F}  \rangle_{2\rightarrow1} = -  \langle \vec{F}
   \rangle_{1\rightarrow2}$ using the same arguments.

   Let     us     assume    now     that     there     is    no     correlation,
   i.e. $P_2(\vec{r}_1,\vec{r}_2) =  P_1(\vec{r}_1)P_1(\vec{r}_2)$, and that the
   fluid  is isotropic  and homogeneous.:  $P_1(\vec{r}_1)=1/V_1$. Choosing  the
   origin  of   positions,  $\vec{x}$,   at  the  center   of  $V_1$,   we  have
   $N_1/V_1=\rho_1(\vec{x})$ (see Sect.~\ref{sec:bbgky}) and we define

   \begin{eqnarray}
     \vec{a}(\vec{x}) &=& \frac{1}{V_1}\int_{V_1}\vec{a}(\vec{r})dV_1, \cr
     \Phi(\vec{x}) &=& \frac{1}{V_1}
     \iint_{V_1,V_2}\phi(|\vec{r}_2-\vec{r}_1|)\rho_2(\vec{r}_2)dV_1 dV_2.
   \end{eqnarray}
   The second law then becomes
   \begin{eqnarray}\label{eq:newton_ideal}
     \frac{\rho_1(\vec{x})m_1 }{V_1}\int_{V_1}\vec{a}(\vec{r})dV_1 &=&
     \frac{\rho_1(\vec{x})m_1 }{V_1}\iint_{V_1, V_2}
     \frac{\vec{F}_{\vec{r}_2\rightarrow \vec{r}_1} }{m} \rho_2(\vec{r}_2) dV_1
     dV_2\cr \rho_1(\vec{x})m_1 \vec{a}(\vec{x}) &\approx& \rho_1(\vec{x})m_1
     \frac{-\vec{\nabla}_{\vec{x}} \Phi(\vec{x})}{m_1},
   \end{eqnarray}
   where one  recognizes now the usual  form of an external  mean force deriving
   from  the  mean  field  potential   $\Phi(\vec{x})$.   In  the  case  of  the
   gravitational force, $\Phi(\vec{x})/m_1$  is independent of $m_1$  and we see
   that the equivalence principle for a pressureless ideal fluid consists simply
   in  simplifying  Eq.~\ref{eq:newton_ideal}  by $\rho_1(\vec{x})m_1$  on  each
   side:
   \begin{equation}
     \vec{a}(\vec{x}) \approx \frac{-\vec{\nabla}_{\vec{x}} \Phi(\vec{x})}{m_1}.
   \end{equation}
   However, this simplification is only  possible in the absence of correlations
   i.e.       as       soon      as       $P_2(\vec{r},\vec{r^\prime})      \neq
   P_1(\vec{r})P_1(\vec{r^\prime})$, the use of the equivalence principle is not
   correct as it will ignore the correlations present in the system.

   \subsection{Non-ideal equation of state}\label{sec:eos}

   We    recall    that    for    a    homogeneous    and    isotropic    fluid,
   $P_2(\vec{r},\vec{r}^\prime)  =  g(|\vec{r}-\vec{r}^\prime|)/V^2$.   In  that
   case (for $N\gg 1$), the interaction energy becomes

   \begin{equation}\label{eq:int_e}
     \langle H_\mathrm{int}  \rangle = \frac{2\pi N^2}{V}  \int_0^R g(r) \phi(r)
     r^2 dr,
   \end{equation}
   with $r=|\vec{r}-\vec{r^\prime}|$ and $V=4\pi R^3/3$. One can use statistical
   mechanics in the  canonical ensemble in thermal equilibrium with  a heat bath
   at  a fixed  temperature $T$  to  define the  EOS.   We refer  the reader  to
   classical textbooks  such as \citet{hansen:2006}  for details. The  energy of
   the system in the absence of external forces is then given by:

   \begin{equation}\label{eq:eos_E}
     E = \frac{Nk_B T}{\gamma-1} +  \frac{2\pi N^2}{V} \int_0^R g(r) \phi(r) r^2
     dr,
   \end{equation}
   with $\gamma=C_P/C_V$ the adiabatic index of  the fluid, defined as the ratio
   of  the  specific  heats  at   constant  pressure  and  volume,  respectively
   ($\gamma=5/3$ for  mono-atomic particles in  their ground state).   The first
   term in Eq. \ref{eq:eos_E} is the perfect gas contribution whereas the second
   term stems from the interactions between particles. The non-ideal pressure is
   given by:

   \begin{equation}\label{eq:eos_P}
     PV = Nk_B T - \frac{2\pi N^2}{3V} \int_0^R g(r) \frac{d\phi(r)}{dr} r^3 dr.
   \end{equation}
   If one identifies  the temperature with the velocity  dispersion of particles
   within the fluid ($k_BT=1/2\langle mv_i^2\rangle$), Eq.~\ref{eq:eos_P} can be
   interpreted as the isotropic form of the Virial theorem. $P=0$ corresponds to
   Virial  equilibrium,  for  which  the  ideal  pressure  is  balanced  by  the
   interaction contribution. $P>0$  corresponds to a fluid  that expands because
   either  the ideal  (kinetic) pressure  dominates the  interaction one  or the
   interactions are  repulsive, $d\phi(r)/dr<0$. In contrast,  $P<0$ corresponds
   to  a  fluid  that  collapses  because the  contribution  due  to  attractive
   interactions $d\phi(r)/dr>0$  dominates the ideal one.  This case corresponds
   for instance  to a phase transition  from gas to liquid,  with the attractive
   interactions  being the  intermolecular  dipole forces,  or to  gravitational
   collapse   in    the   case   of   the    gravitational   force   \citep[see,
     e.g.][]{devega:2002}.

   Strictly speaking, a self-interacting fluid should always be described with a
   non-ideal  EOS.  Nevertheless,  an  ideal approximation  is  possible if  the
   interaction term is  negligible in Eq.~\ref{eq:eos_P}.  Indeed,  in that case
   $g(r)\approx 1$, so that, with $\rho=N/V$, the condition becomes
   \begin{equation}
     \rho k_B T \gg \frac{2\pi \rho^2}{3} \int_0^R \frac{d\phi(r)}{dr} r^3 dr.
   \end{equation}
   For a potential of the form  $\phi = -k/r$ (with $k>0$ for  the gravitational
   force or attractive Coulomb interactions)
   \begin{eqnarray}\label{eq:cond_ideal}
     \rho k_B T \gg  \pi \rho^2 k R^2 /3 \Rightarrow  R \ll \sqrt{\frac{k_B T}{k
         \rho}}.
   \end{eqnarray}
   In this case, one can recognize  the characteristic Debye screening length in
   the case  of Coulomb  interactions or  the Jeans  length in  the case  of the
   gravitational force. The condition  in Eq.~\ref{eq:cond_ideal} states that an
   ideal  approximation in  the treatment  of the  fluid under  consideration is
   acceptable if (and only if) the  volumes considered are sufficiently small so
   that the Debye  length or the Jeans length is  well-resolved.  Whereas such a
   condition is usually  fulfilled in numerical simulations  of star formation,
   and is known as the  'Truelove condition' \citep[see][]{truelove:1997}, it is
   not easily done  for large-scale numerical simulations of  galaxies or larger
   structures. In these cases,  it is mandatory to use a  non-ideal EOS, thus to
   include  in  the  calculations   the  terms  involving  the  pair-correlation
   function.

   \subsection{Non-ideal stress tensor}

   More generally,  one can  express the above  theory in the  form of  a stress
   tensor, following the pioneering work of \citet{kirkwood:1949}. The Newtonian
   expression for the stress tensor of a system of $N$ particles  of mass $m$,
   is given by

   \begin{equation}
     \sigma_{ij}  = -\left(P+\left(\frac{2}{3}\eta-\chi\right)\sum_k  \partial_k
     u_k\right)\delta_{ij}+\eta \left( \partial_i u_j +\partial_j u_i \right),
   \end{equation}
   with $P$ the  non-ideal pressure given in Sect.~\ref{sec:eos}  and $\eta$ and
   $\chi$ the coefficients of shear and bulk viscosity that can be computed as
   \begin{eqnarray}
     \eta  &=& \frac{1}{2}  \frac{N}{V}  m D  + \frac{\pi}{15D}  \frac{N^2}{V^2}
     \int_0^R  \psi_2(r)   g(r)  \frac{d\phi(r)}{dr}   r^3  dr,  \cr   \chi  &=&
     \frac{1}{3}\frac{N}{V}    m    D     +    \frac{\pi}{9D}    \frac{N^2}{V^2}
     \int_0^R\psi_0(r) g(r) \frac{d\phi(r)}{dr} r^3 dr,
   \end{eqnarray}
   with $D$ the  self-diffusion coefficient in the fluid (from  Fick's law).  As
   defined  in \citet{kirkwood:1949},  $\psi_0(r)$  is the  surface harmonic  of
   zeroth order arising from the dilatation component of the rate of strain, and
   $\psi_2(r)$  is the  surface  harmonic  of order  2  arising  from the  shear
   component. They obey the following differential equations
   \begin{eqnarray}\label{eq:psi}
     \frac{d}{dr}\left(     r^2      g(r)\frac{d     \psi_2(r)}{dr}     \right)-
     6\psi_2(r)g(r)&=&   r^3\frac{d   g(r)}{dr}   \cr   \frac{d}{dr}\left(   r^2
     g(r)\frac{d \psi_0(r)}{dr} \right)&=& r^3 \frac{d g(r)}{dr}.
   \end{eqnarray}
   As for the  non-ideal pressure, the non-ideal coefficients of  shear and bulk
   viscosity contain two  contributions: an ideal (Brownian) one arising  from
   the momentum 
   transfer between  colliding particles and  a non-ideal part arising  from the
   direct transfer  by interaction  forces. The  second part  strongly dominates
   when interactions are present in the fluid. As for the non-ideal pressure, it
   is also  crucial to take  into account interactions  in the stress  tensor as
   soon as the Jeans length is not resolved.

   \subsection{Thermodynamic limit}

   We can  take the  thermodynamic limit  by imposing  $(N,V)\rightarrow \infty$
   while keeping  $N/V=\rho$ constant. Defining  $\rho_m e$ the  internal energy
   per unit volume (with $\rho_m=\rho m$):

   \begin{equation}\label{eq:eos_p}
     P    =    \rho    k_B    T   -    \frac{2\pi    \rho^2}{3}    \int_0^\infty
     g(r,\rho,T)\frac{d\phi(r)}{dr} r^3 dr,
   \end{equation}

   \begin{equation}\label{eq:eos_e}
     \rho_m  e  =  \frac{\rho  k_B  T}{\gamma-1}  +  2\pi  \rho^2  \int_0^\infty
     g(r,\rho,T)\phi(r) r^2 dr,
   \end{equation}
   where we have made the dependence  of the radial distribution function to the
   external variables $\rho$  and $T$ explicit. Determining the  EOS now reduces
   to the determination  of $g(r,\rho,T)$, which can be done  by three different
   technics  in  the  case  of  liquids:  calculation  of  the  Ornstein-Zernike
   equation, N-body Monte-Carlo or molecular dynamic simulations, or experiments
   by X-ray  or neutron  diffraction \citep[see][]{aslangul:2006}.  We emphasize
   that we implicitly  assume here that an equilibrium can  be reached, allowing
   to define a radial distribution function that does not depend on time and can
   be determined  from external variables  ($\rho$ and $T$).  Strictly speaking,
   this might not be the case  for self-gravitating systems that are collapsing,
   hence are not in equilibrium. This  assumption then assumes that some sort of
   feedback stabilizes  small scales allowing for  the definition of an  EOS for
   the scales that are not resolved  and described by statistical mechanics. One
   can  relax this  assumption by  using the  dynamical equations  in the  BBGKY
   hierarchy describing the evolution of the correlations. However, as discussed
   in  \citet{davis:1977},  one  needs  to  find  a  closure  relation  for  the
   hierarchy, expressing the last N-point correlation functions as a function of
   the others. This closure is likely possible only at very high $N$ in order to
   describe  accurately  the  structures  that  are  observed  in  the  Universe
   \citep[see][]{balian:1989a,balian:1989b},  which would  result  in a  complex
   system of  dynamical equations describing  the evolution of  correlations. We
   therefore use here  a pragmatic approach (similarly to the  approach used for
   liquids in physical chemistry), assuming that an equilibrium, or more exactly
   a stationary state exists at small scales, and assuming that we can determine
   the   two-point  correlation   function   either   by  explicitly   resolving
   inhomogeneities  in small-scale  numerical simulations  (with e.g.  molecular
   dynamics for liquids or e.g. N-body simulations for self-gravitating systems)
   or by using astrophysical observations.

   An important point  is that the thermodynamic limit  in Eq.~\ref{eq:eos_e} is
   well-defined only  if we get  an extensive  definition of the  total internal
   energy $E$  (or intensive  for the  internal energy  density per  unit volume
   $\rho_m e$),  i.e. the  integral in  the expression  must converge  when $N,V
   \rightarrow \infty$.  Since  $\lim_{r\rightarrow \infty} g(r) =  1$, it means
   that the  potential from which  the force  derives must decrease  faster than
   $r^{-4}$  at infinity.   This is  the case  for the  Lennard-Jones potential,
   classically used to model inter-molecular forces in liquid water or molecular
   liquids. An interpretation of this requirement is that the interaction energy
   in Eq.~\ref{eq:eos_E} is actually not  proportional to $N^2$ but to $N\times
   N_\mathrm{neighbors}$,  with $N_\mathrm{neighbors}$  the number  of particles
   involved in  the short-range interactions. Therefore,  the interaction energy
   is  proportional  to $N$  and  the  definition of  $E$  is  extensive in  the
   thermodynamic  limit.   If  the  interactions are  long-range,  however,  the
   interaction energy is  really proportional to $N^2$. In that  case, the total
   energy $E$ is  not extensive and we cannot use  the thermodynamic limit. This
   problem is  usually thought to  prevent the  use of statistical  mechanics to
   describe systems characterized  by a long-range force, like  gravity. We will
   show in Sect.~\ref{sec:grav-eos}  how to get a correct treatment  in the case
   of the gravitational force.

   \section{Non-ideal self-gravity} \label{sec:grav-eos}

   \subsection{Hierarchical Virial theorem}\label{sec:virial_hier}

   As we have seen in  Sect.~\ref{sec:eos}, characteristic length scales such as
   the Debye  length $\lambda_D$ or Jeans  length $\lambda_J$ can be  defined to
   check  out whether  or not  non-ideal (interaction)  contributions should  be
   taken into  account.  In  the case  of the  electrostatic force,  the Coulomb
   potential is  split into a  short-range and  a long-range component,  $\phi =
   \exp(-r/\lambda_D)\phi   +  (1-\exp(-r/\lambda_D))\phi$.    The  first   part
   corresponds  to  the  screened  Coulomb potential,  due  to  the  polarizable
   electron background,  and the residual  corresponds to the  departure between
   the  Coulomb and  the  screened  potential \citep[e.g.][]{chabrier:1990}.  We
   propose to decompose the gravitational force in a similar manner.

   Hydrogen  atoms  get  gravitationally  bounded  e.g. in  the  interior  of  a
   star. Using  a classical  polytropic stellar  model, we  can define  a Virial
   theorem that  balance the internal energy  in the star and  the gravitational
   interaction  energy  (see  e.g.   Sect.~\ref{sec:appli_virial}).  If  now  we
   consider a cluster of interacting  polytropic stars, in a first approximation
   we can consider each star as isolated  and use the previous Virial theorem to
   get  their internal  structure and  then describe  their collective  dynamics
   assuming they are  point mass (i.e. we neglect  tidal interactions). However,
   one can  easily see that we  should not integrate the  gravitational force at
   infinity when using  the Virial theorem in a single  star, otherwise we would
   include into the internal stellar structures some gravitational interactions
   that contribute to the cluster dynamics. A solution to this issue is to decompose the
   gravitational potential  into a  near-field and  far-field component  using a
   length scale $\lambda_0$ larger than the size of a star
   \begin{equation}
     \phi = e^{-r/\lambda_0}\phi + \left(1-e^{-r/\lambda_0}\right)\phi
   \end{equation}
   and we apply the Virial theorem inside  a star using the near-field $\phi_0 =
   \exp(-r/\lambda_0)\phi$. We emphasize that the  use of an exponential damping
   is arbitrary.  Another possibility is  to use an  erf function as  often done
   with  the Coulomb  interaction in  physical chemistry.   We can  then proceed
   recursively  on  the  residual  part.  We  use  now  a  second  length  scale
   $\lambda_1$ in order  to define a near-field $\phi_1$ for  the Virial theorem
   describing  the dynamics  inside  the  stellar cluster  and  a far-field  for
   e.g.  the dynamics  of  a collection  of stellar  clusters  inside a  galaxy.
   $\phi_1$           is           then           given           by           $
   \exp(-r/\lambda_1)(1-\exp(-r/\lambda_0))\phi$. One can continue to define the
   gravitational energy  contributing to  the dynamics  of the  stellar clusters
   inside  the galaxy  and then  the dynamics  of the  galaxies inside  a galaxy
   cluster  and so  on.   The hierarchical  decomposition  of the  gravitational
   potential       with      a       collection      of       length      scales
   $\lambda_0$,... $\lambda_i$... is then given by
   \begin{eqnarray}
     \phi     &=&     \sum_{i=0}^{\infty}      \phi_i     \cr     \phi_i     &=&
     e^{-r/\lambda_i}\prod_{j=0}^{i-1}\left(1-e^{-r/\lambda_j}\right) \phi
   \end{eqnarray}
   Note that this expansion is not without recalling the BBGKY hierarchy for the
   N-body problem  in statistical physics, where  the potential is split  into a
   (short-range) inter-particle pair potential and a (long-range) external-field
   potential \citep[see, e.g.,][]{chavanis:2013}. The series of length scales is
   a-priori arbitrary and for a  purely self-gravitating collapsing fluid, there
   is  no  particular  length  scale  that  would  physically  make  sense.  The
   characteristic  length  scales that  should  be  used  are likely  linked  to
   feedback and support processes that  define the possible Virial equilibria at
   different scales:  AGN feedback for  galaxy clusters, rotational  support for
   galaxies,  stellar  feedback for  molecular  clouds  and star  clusters,  and
   pressure  support  for stars.  Once  such  an  stationary state  is  reached,
   e.g.  inside  a  star,  one  can account  for  the  velocity  dispersion  and
   gravitational energy at this scale  as internal degrees of freedom (similarly
   to an EOS) when going to a  larger scale. If one neglects tidal effects, this
   allows to decouple the different scales based on feedback processes.

   We can  now use  the Virial
   theorem  defined in  Sect.~\ref{sec:virial}  to link  the dynamic  properties
   $\vec{v}_i, I_i$  of objects of mass  $m_i$ at each scale  $\lambda_i$ of the
   hierarchy to the corresponding gravitational interaction potential $\phi_i$:
   \begin{equation}\label{eq:virial_hier}
    - \langle  m_i  v_i^2  \rangle  +\langle  d^2I_i/dt^2  \rangle/2  =  \langle
    \vec{F}_i\cdot \vec{r}\rangle,
   \end{equation}
   in  which we  can replace  kinetic energy  by thermal  energy for  $i=0$.  As
   mentioned   above,  we   emphasize   that  this   hierarchy  neglects   tidal
   interactions.  A possible  extension would  be to  also decompose  the energy
   going into tides, if they are not negligible at a given scale. This is likely
   to be  the case  at the scale  of galaxies  \citep[see, e.g.][]{barnes:1992}.
   For  the  time  being,  however,  we will  assume  that  these  contributions
   represent a small correction to the total gravitational energy.

   Since each term $\langle \vec{F}_i\cdot \vec{r}\rangle $ is computed with the
   short-range potential $\phi_i$, damped by a factor $e^{-r/\lambda_i}$, we can
   now  safely   take  the  thermodynamic  limit   in  Eq.~\ref{eq:virial_hier},
   i.e.  $N,V  \rightarrow  \infty$,  since   the  integral  converges.   For  a
   homogeneous and isotropic correlated fluid, this term is given by
   \begin{equation}
     \langle \vec{F}_i\cdot  \vec{r}\rangle =  2\pi \rho^2  \int_{0}^\infty g(r)
     \frac{d\phi_i(r)}{dr} r^3 dr.
   \end{equation}
   Similarly, the total  mechanical energy of the system at  a scale $\lambda_i$
   is given by
   \begin{equation}
     E_{m,i} = \langle m_i v_i^2/2 \rangle + \langle H_\mathrm{int,i} \rangle,
   \end{equation}
   where we can take the thermodynamic limit for the interaction energy:
   \begin{equation}
     \langle  H_\mathrm{int,i}  \rangle  =   2\pi  \rho^2  \int_{0}^\infty  g(r)
     \phi_i(r) r^2 dr
   \end{equation}
   We will see in  the next section that a similar decomposition  can be used to
   define non-ideal self-gravitating hydrodynamics in numerical simulations.

   \subsection{Non-ideal self-gravitating hydrodynamics}
   \label{sec:navier-stockes}

   A standard  strategy to  obtain a large-scale  hydrodynamic model  of complex
   multi-phasic  systems relies  on the  homogenization procedure,  by averaging
   volumes of characteristic size  $\lambda$ \citep{whitaker:1998}. One can then
   decompose  the  interaction  potential  into a  near-field  and  a  far-field
   contribution,  using  this  characteristic   scale.   However,  in  numerical
   simulations,  as soon  as  the grid  size  (or the  kernel  extension in  SPH
   methods) is such that $\Delta x < \lambda$, inhomogeneities at scales smaller
   than $\lambda$ have no physical meaning in such a homogenized  model that  is
   only valid at scales larger than $\lambda$.

   In  numerical simulations,  it is  then natural  to use  the grid  resolution
   $\Delta  x$  as   the  characteristic  length  to   split  the  gravitational
   interaction
   \begin{equation}
     \phi(r) = e^{-r/\Delta x}\phi(r) + \left(1-e^{-r/\Delta x}\right)\phi(r)
   \end{equation}
   We  can  then  define  a near-field  $\phi_\mathrm{int}(r)  =  \exp(-r/\Delta
   x)\phi(r)$ to handle  the non-ideal effects in the EOS  inside the simulation
   control volumes,  and a  far-field $\phi_\mathrm{ext}(r)  = (1-\exp(-r/\Delta
   x))\phi(r)$  to define  the  external force  between  the simulation  control
   volumes.  Following  \citet{irving:1950}, the Navier-Stokes equations  for a
   non-ideal  fluid can  be derived  from the  Liouville equation  in the  BBGKY
   hierarchy and in the case of a self-gravitating fluid are given by
   \begin{eqnarray}\label{eq:navier-stockes}
     \frac{\partial    \rho_m}{\partial   t}    +   \vec{\nabla}\left(    \rho_m
     \vec{u}\right)  &=& 0,  \cr  \frac{\partial \rho_m  \vec{u}}{\partial t}  +
     \vec{\nabla}\left(\rho_m  \vec{u}\otimes\vec{u} +  \tens{\sigma}\right) &=&
     \vec{F}_\mathrm{ext},\cr   \frac{\partial    \rho_m   E}{\partial    t}   +
     \vec{\nabla}\left(\rho_m \vec{u} E  + \tens{\sigma}\cdot\vec{u} \right) &=&
     \vec{F}_\mathrm{ext}\cdot \vec{u},
   \end{eqnarray}
   using the following relations for $E$ and $\tens{\sigma}$
   \begin{eqnarray}
     E      &=&     e      +      \frac{1}{2}\vec{u}^2,     \cr      \sigma_{ij}
     &=&-\left(P+\left(\frac{2}{3}\eta-\chi\right)\sum_k              \partial_k
     u_k\right)\delta_{ij}+\eta \left( \partial_i u_j +\partial_j u_i \right),
   \end{eqnarray}
   in which the non-ideal (internal) quantities are given by
   \begin{eqnarray}
     P    &=&   \rho    k_B    T   -    \frac{2\pi   \rho^2}{3}    \int_0^\infty
     g(r)\frac{d\phi_\mathrm{int}(r)}{dr} r^3 dr\cr \rho_m  e &=& \frac{\rho k_B
       T}{\gamma-1} +  2\pi \rho^2  \int_0^\infty g(r)  \phi_\mathrm{int}(r) r^2
     dr\cr \eta  &=& \frac{1}{2} \rho_m D  + \frac{\pi}{15D}\rho^2 \int_0^\infty
     \psi_2(r)   g(r)   \frac{d\phi_\mathrm{int}(r)}{dr}   r^3   dr   \cr   \chi
     &=&\frac{1}{3}  \rho_m  D  + \frac{\pi}{9D}\rho^2  \int_0^\infty  \psi_0(r)
     g(r)\frac{d\phi_\mathrm{int}(r)}{dr}       r^3       dr      \cr       {\rm
       with}\,\,\phi_\mathrm{int}(r) &=& -\frac{Gm^2}{r}e^{-r/\Delta x},
   \label{eos}
   \end{eqnarray}
   with  $\psi_2(r)$ and  $\psi_0(r)$ given  by Eq.~\ref{eq:psi}.   The external
   force  is  defined  by  an  integral form  on  the  whole  simulation  domain
   $V_\mathrm{sim}$
   \begin{eqnarray}
     \vec{F}_\mathrm{ext}(\vec{x})                                           &=&
     -\rho(\vec{x})\vec{\nabla}_{\vec{x}}(\Phi_\mathrm{ext}(\vec{x})),\cr
     \Phi_\mathrm{ext}(\vec{x})                                              &=&
     \int_{V_\mathrm{sim}}\phi_\mathrm{ext}(|\vec{x}-\vec{x}^\prime|)\rho(\vec{x}^\prime)
     dV_\mathrm{sim},\cr      {\rm     with}\,\,\,\,\,      \phi_\mathrm{ext}(r)
     &=&-\frac{Gm^2}{r}(1-e^{-r/\Delta x})
   \end{eqnarray}
   As for the Virial  theorem in the previous section, all  the integrals in the
   EOS     quantities    are     defined    with     the    damped     potential
   $\phi_\mathrm{int}(r)=e^{-r/\Delta x}\phi(r)$. Hence, all these integrals are
   well-defined and  converge in  the thermodynamic limit.   When $\Delta  x \ll
   \lambda_J$  and notably  in  the  limit $\Delta  x  \rightarrow  0$, we  have
   $\phi_\mathrm{int}(r)  \rightarrow 0$  and $\phi_\mathrm{ext}  (r)\rightarrow
   \phi(r)$, which means  that all the gravitational force  is entirely resolved
   externally  and  the  non-ideal  contributions  to the  fluid  EOS  given  by
   Eqn.(\ref{eos}) are negligible: we recover then the classical (ideal) form of
   the Navier-Stokes equations for a self-gravitating fluid. As soon as $\Delta
   x  > \lambda_J$,  however,  the non-ideal  contributions to  the  EOS due  to
   interactions  within the  control  volumes  must be  included  for a  correct
   calculation of the gravitational energy of the whole system.

   \begin{figure}
   \begin{centering}
     \includegraphics[width=\linewidth]{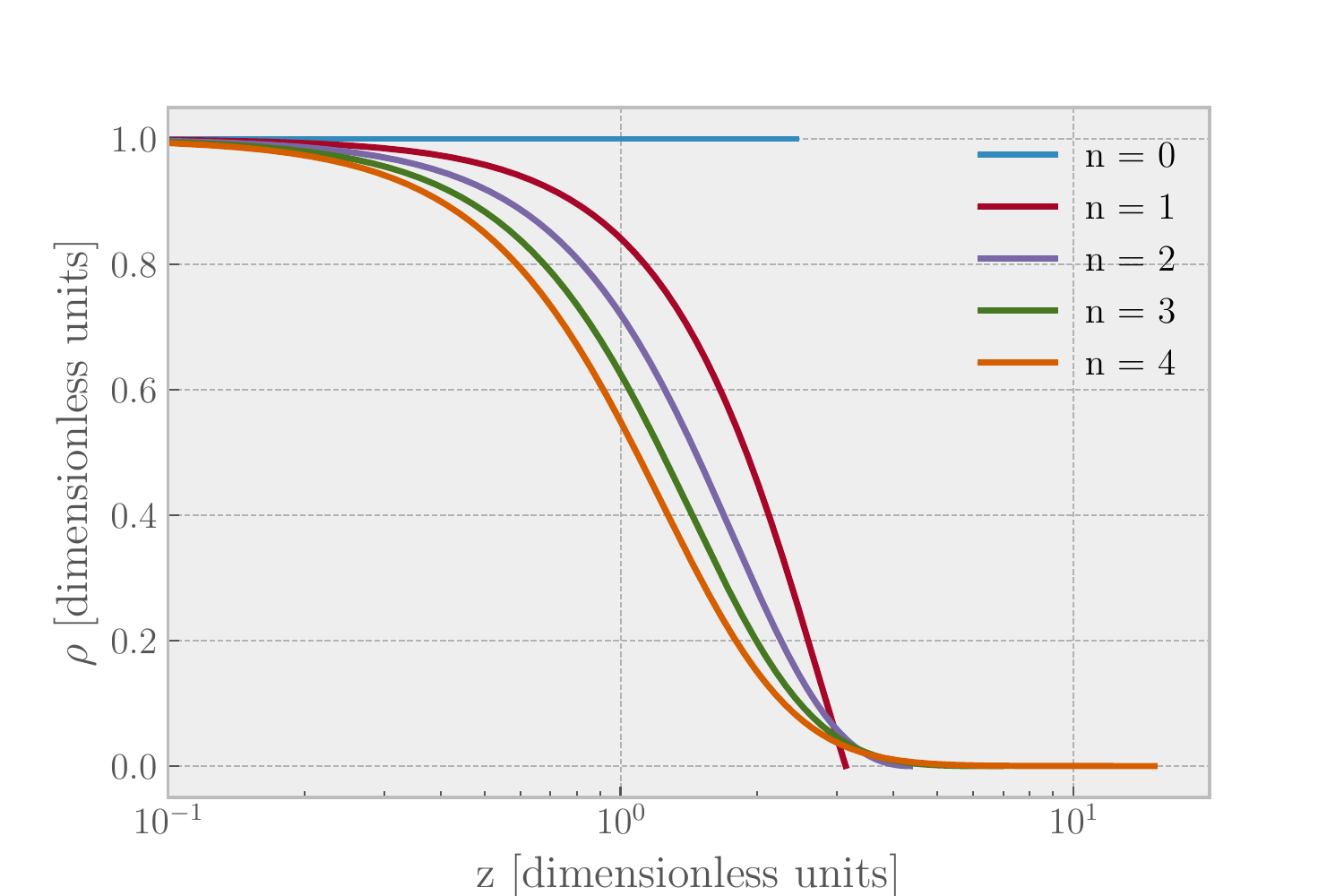}
     \includegraphics[width=\linewidth]{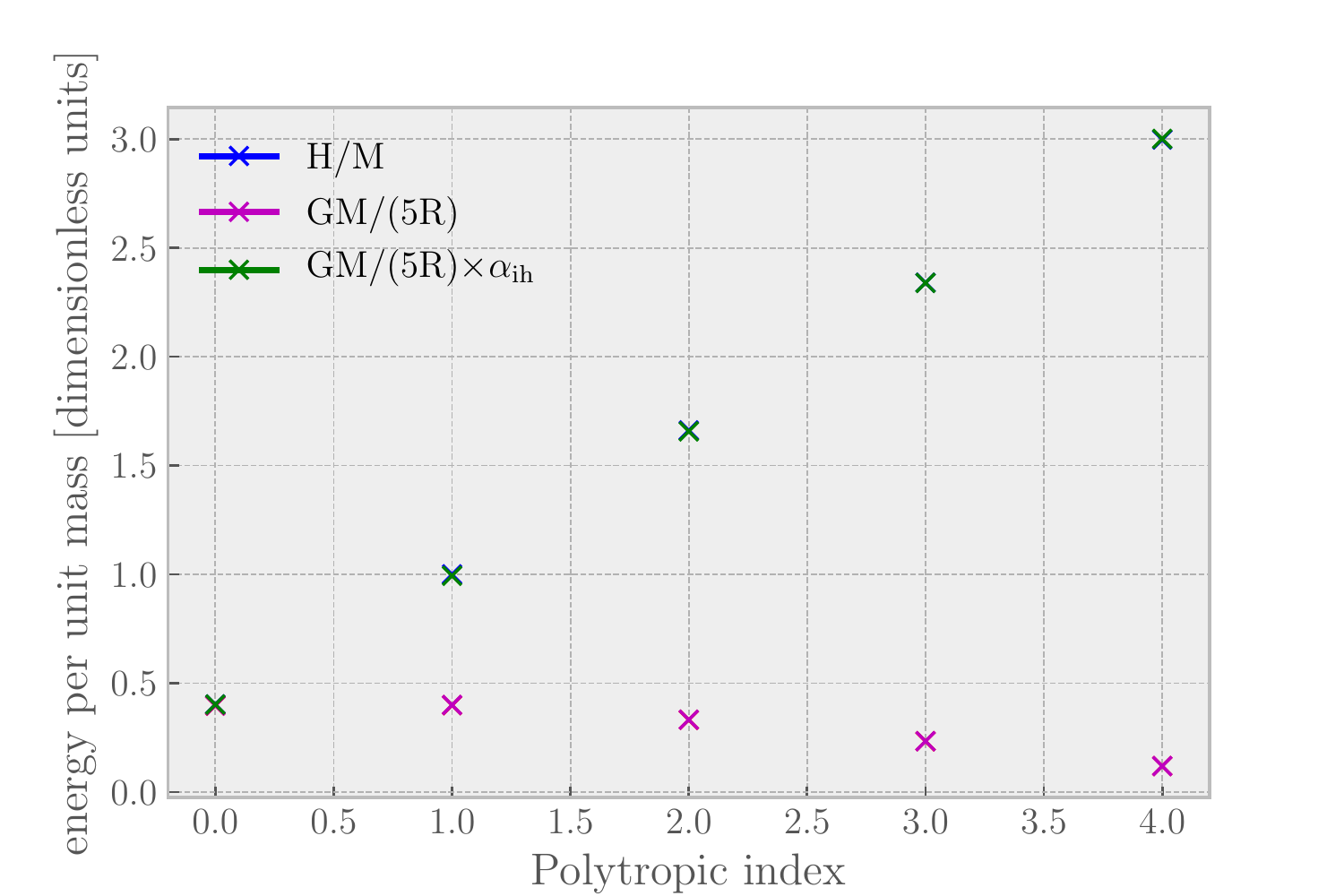}
     \caption{Top: density  profiles solutions  to the Lane-Emden  equation with
       different polytropic indexes  $n$.  Bottom: total enthalpy  per unit mass
       compared  to  gravitational interaction  energy  per  unit mass  assuming
       either a  homogeneous density profile or  accounting for inhomogeneities.
       Blue and green crosses are indistinguishable.} \label{fig:polytrops}
   \end{centering}
   \end{figure}

   \section{Applications}\label{sec:appli}

   \subsection{Impact of inhomogeneities on the Virial theorem}
   \label{sec:appli_virial}

   The Virial  theorem is usually applied  in astrophysics in a  simplified way.
   We define the total  mass $M=N m$, and the average  total energy $\langle E/M
   \rangle$,  kinetic  energy  $\langle E_\mathrm{kin}/M\rangle$  and  potential
   energy $\langle E_p/M  \rangle$ per unit mass.  The  Virial theorem, assuming
   that  the  particles  are  gravitationally  bound  in  a  sphere  of  radius
   $r_\mathrm{bound}$, is usually evaluated with these quantities per unit mass
   \begin{eqnarray}
     \langle  E/M \rangle  &=& \langle  E_\mathrm{kin}/M \rangle  +\langle E_p/M
     \rangle, \cr  -2\langle E_\mathrm{kin}/M \rangle &=&  \langle E_p/M \rangle
     \approx -\frac{G M}{r_\mathrm{bound} }.
   \end{eqnarray}
   One can  see that the  use of a radius  $r_\mathrm{bound}$ is similar  to the
   decomposition  into  near-   and  far-fields  that  we   have  introduced  in
   Sect.~\ref{sec:virial_hier}. It defines a scale up to which the gravitational
   energy is considered  to contribute to the Virial balance,  while the rest is
   implicitly assumed to contribute to  larger-scale dynamics. However, one also
   sees that this  form of the Virial theorem contains  no information about the
   sub-structures present within the bound volume:  only the total mass and size
   of  the system  enters into  the Virial  equation. This  latter thus  ignores
   non-ideal effects due to correlations induced by these sub-structures.

   In order to show the importance  of inhomogeneities in the Virial theorem, we
   will  apply  it, as  a  simple  semi-analytical example,  to  the  case of  a
   polytropic  stellar structure.  In such  a  case, we  can explicitly  compute
   $P_2(\vec{r},\vec{r}^\prime)$ by computing semi-analytically an inhomogeneous
   density field  and compare the  result with  a homogeneous assumption  on the
   Virial  theorem. A  polytropic stellar  structure  can be  calculated by  the
   Lane-Emden equation
   \begin{equation}
     \frac{1}{z^2}\frac{d}{dz}\left( z^2 \frac{dw}{dz}\right) + w^n = 0,
   \end{equation}
   with $\rho  = \rho_c w^n$, $P=K\rho^\gamma$,  $n=1/(\gamma-1)$ the polytropic
   index,  $z=Ar$,   and  $A^2=(4\pi  G/(n+1))\rho_c^{1-1/n}/K$.  We   choose  a
   dimensionless unit system in which  $\rho_c=K=G=1$. By solving the Lane-Emden
   equation for different polytropic indexes $n<5$, we can compute inhomogeneous
   density  profiles  that have  a  finite  radius $R$  (see  the  top panel  of
   Fig.~\ref{fig:polytrops}). For a polytrope, the following Virial theorem (per
   unit mass) reads:
   \begin{eqnarray}
     H/M &= 2H_\mathrm{int}/M, \cr H &= \int_V \frac{\gamma}{\gamma-1}P dV,\cr M
     &= \int_V \rho dV,
   \end{eqnarray}
   where $H$ is  the total enthalpy, $M$ the total  mass and $H_\mathrm{int}$ is
   given by Eq.~\ref{eq:hint_ideal} with the zero of potential energy defined at
   the   surface  of   the  star.   For   a  homogeneous   density  profile   in
   Eq.~\ref{eq:hint_ideal}, with $\rho =M/(4\pi R^3/3)$, we get

   \begin{equation}\label{eq:ehomo}
     \frac{H}{M} = \frac{GM}{5R}.
   \end{equation}

   For the  inhomogeneous case,  as illustrated  in Fig.~\ref{fig:distribution},
   there is two possibilities for the description of the inhomogeneities. In the
   first case  (with $V_1$),  these latter  are resolved  (for instance  in high
   resolution simulations or observations) so  that $\rho(r)$ is no constant. In
   the opposite  case (with $V_2$),  $\rho(r)$ is constant and  the (unresolved)
   inhomogeneities are included  in the pair correlation function  $g(r)$ in our
   formalism.  Finding out the impact of ignoring the inhomogeneities thus leads
   to two different  possibilities. For the ‘resolved’ case,  it means comparing
   the computation with  $\rho(r)$ not constant with the one  in which $\rho(r)$
   is replaced by its mean value.  For the ‘unresolved’ case, it means comparing
   the computation with $g(r)\ne 1$ with the one assuming an uncorrelated fluid,
   $g(r)= 1$.  The semi-analytical polytropic model allows to easily perform the
   first  (‘resolved’)  test.  Indeed,  since  we  can  compute  explicitly  the
   inhomogeneous profile $\rho(r)$ and all the necessary integral quantities, we
   can compare this exact solution with  the one replacing $\rho(r)$ by its mean
   value.

   Let us  now assume that  an observer has only  access to the  total enthalpy,
   mass and radius of different objects  and does not know whether these objects
   have substructures  or not. The naive  assumption, as explained above,  is to
   ignore the possible inhomogeneities and  assume that the density is constant,
   $\rho  =  M/(4\pi R^3/3)$,  hence  using  Eq.~\ref{eq:ehomo} for  the  Virial
   equation.   Such an  assumption, however,  is only  correct for  a polytropic
   index $n=0$.  As  shown in the bottom panel  of Fig.~\ref{fig:polytrops}, for
   profiles $n\ne 0$, the error on the  total energy is significative and can be
   up  to a  factor 25  for $n=4$.  Therefore, the  assumption of  a homogeneous
   profile within the  structures will lead to a 'missing  mass problem'.  For a
   polytropic  structure,  however,  we   can  calculate  semi-analytically  the
   inhomogeneous density profile, and thus calculate exactly the contribution of
   inhomogeneities to the total energy.  We characterize this contribution by an
   inhomogeneous   amplification  factor   $\alpha_\mathrm{ih}$  which   can  be
   calculated exactly:

   \begin{eqnarray}
     \alpha_\mathrm{ih}   &=&   \frac{    N^2   \iint_{V,V}   P_2(\vec{r},
       \vec{r}^\prime)    \phi(|\vec{r}-\vec{r}^\prime|)    dV    dV^\prime}{N^2
       \iint_{V,V}  1/(4\pi R^3/3)^2  \phi(|\vec{r} -  \vec{r}^\prime|) dV
       dV^\prime}\cr    &=&\frac{\iint_{V,V}     \rho(r)    \rho(r^\prime)
       (1/|\vec{r}               -               \vec{r}^\prime|               -
       1/||\vec{R}-\vec{r}^\prime|)dVdV^\prime}{\iint_{V,V}      (M/(4\pi
       R^3/3))^2(1/|\vec{r}  - \vec{r}^\prime|  - 1/|\vec{R}  - \vec{r}^\prime|)
       dVdV^\prime},
   \end{eqnarray}
   with  $\vec{R}$ a  vector  on a  sphere  of  radius $R$  (the  radius of  the
   polytropic structure), used to define the zero of the gravitational potential
   at  the  stellar surface,  and  $V=4\pi  R^3/3$.   In our  calculation,  $N^2
   P_2(\vec{r},\vec{r}^\prime)     =    N^2P_1(\vec{r})P_1(\vec{r}^\prime)     =
   \rho(\vec{r}) \rho(\vec{r}^\prime)$ for the inhomogeneous structure while the
   homogeneous  calculation corresponds  to  $N^2 P_2(\vec{r},\vec{r}^\prime)  =
   N^2P_1(\vec{r})P_1(\vec{r}^\prime) =  N^2/V^2$, with  $P_1(\vec{r})=1/V$.  We
   can now correct Eq.~\ref{eq:ehomo} to account for inhomogeneities
   \begin{equation}\label{eq:ehomo2}
     \frac{H}{M} = \frac{GM}{5R} \alpha_\mathrm{ih}.
   \end{equation}
   As shown  in the bottom  panel of Fig.~\ref{fig:polytrops}, when  taking into
   account this factor, we recover exactly  the correct enthalpy and the correct
   gravitational  interaction energy  for  the  different polytropic  structures
   (note that the  green and blue crosses in the  figure are indistinguishable).
   This simple  example demonstrates  that it  is mandatory  to account  for the
   contribution of sub-structures to  correctly evaluate interaction energies in
   astrophysical structures.

   \subsection{Gravitational viscous regime and the rotation curve of galaxies}

   Another  consequence  of  sub-structures   and  interactions  is  a  possible
   transition to  a viscous regime  when interactions dominate at  small scales.
   Heuristically, one  can expect the rotation  curve of a fluid  to be strongly
   impacted  by viscous  stresses. Indeed,  for a  large viscosity  the rotation
   tends to  the rotation  of a Taylor-Couette  flow with  $u_\theta \rightarrow
   \Omega r$ in cylindrical coordinates.

   We can  explore different regimes  by performing  a Hilbert expansion  with a
   small parameter $\epsilon$ that characterizes the regime of the flow, i.e. we
   can take $\epsilon$ equal to the inverse  of the Reynolds number Re such that
   $\epsilon \rightarrow  0$ in the  inertial limit.  We assume a  stationary 2D
   stellar flow  in cylindrical  coordinates using either  stellar hydrodynamics
   \citep[see][]{binney:1987,burkert:1988} or  gas hydrodynamics in the  HI disk
   with    the   non-ideal    momentum   evolution    equation   presented    in
   Sect.~\ref{sec:navier-stockes}. In the case of stellar hydrodynamics, one can
   replace  $k_BT$ in  Eq.~\ref{eq:navier-stockes}  by  the velocity  dispersion
   tensor, as  done in  the Jeans  equation, while the  interaction part  in the
   non-ideal term remains  unchanged. We develop all the  variables with respect
   to $\epsilon$
   \begin{eqnarray}
     \rho_m(r,\theta)  &=&  \rho_m^0(r)   +  \epsilon  \rho_m^1(r,\theta)+...\cr
     u_r(r,\theta) &=& \epsilon u_r^1(r,\theta)  + ...\cr u_\theta(r,\theta) &=&
     u_\theta^0(r) + \epsilon u_\theta^1(r,\theta)+...\cr P(r,\theta) &=& P^0(r)
     +   \epsilon    P^1(r,\theta)+...\cr   \vec{F}_\mathrm{ext}(r,\theta)   &=&
     \vec{F}_\mathrm{ext}^0(r)                     +                    \epsilon
     \vec{F}_\mathrm{ext}^1(r,\theta)+...\cr \eta &=& \eta^0 + \epsilon \eta^1 +
     ...
   \end{eqnarray}
   where we have made the hypothesis that the zeroth order is axi-symmetric with
   $u_r^0=0$, and verifies the condition $\vec{\nabla}(\vec{u^0}) = 0$.

   We can now  explore the inertial regime by assuming  the following dependence
   of the shear  viscosity and pressure, $\eta^0 \approx 0$  and $P^0(r) \approx
   0$ in  the ideal  regime, thus  with no  contribution from  interactions. The
   zeroth     order     of     the     non-ideal     Navier-Stokes     equation
   (Eq. \ref{eq:navier-stockes}) then gives
   \begin{eqnarray}
     0^{\mathrm{th}},r\mathrm{-component:}&&     \frac{-(u_\theta^0)^2}{r}     =
     \frac{1}{\rho_m^0}F_\mathrm{ext,r}^0\cr
     0^{\mathrm{th}},\theta\mathrm{-component:}&& 0=0,
   \end{eqnarray}
   which  gives  the  classical   velocity  profile  $u_\theta^0  =\sqrt{GM/r}$,
   decreasing as $r^{-1/2}$ at large distance.

   In the viscous  regime dominated by the non-ideal contribution,  we now have
   $\eta^0 \neq 0$  and $P^0(r) \neq 0$ because of  the presence of interactions
   and  bound  structures. The  zeroth  order  of the  non-ideal  Navier-Stokes
   equation then gives
   \begin{eqnarray}
     0^{\mathrm{th}},r\mathrm{-component:} && \frac{-(u_\theta^0)^2}{r} = -
     \frac{1}{\rho_m^0}\frac{\partial         P^0}{\partial         r}         +
     \frac{1}{\rho_m^0}F_\mathrm{ext,r}^0\cr
     0^{\mathrm{th}},\theta\mathrm{-component:} && 0=  \eta^0 \left( \frac{1}{r}
     \frac{\partial} {\partial r} \left(  r \frac{\partial u^0_\theta} {\partial
       r} \right) - \frac{u_\theta^0} {r^2} \right).
   \end{eqnarray}
   We  can  get  the  velocity   profile  from  the  $\theta$-component  of  the
   Navier-Stokes equation.   Under appropriate  boundary conditions,  the fluid
   can exhibit a Taylor-Couette flow in  the viscous limit: $u_\theta^0 = \Omega
   r$. The associated centrifugal force in the $r$-component is then compensated
   by  pressure  gradients.  We  recall  that  the non-ideal  pressure  contains
   attractive  gravitational  interactions at  small-scale  here,  hence it  can
   compensate the centrifugal outward acceleration.

\begin{figure}
   \begin{centering}
     \includegraphics[width=0.48\linewidth]{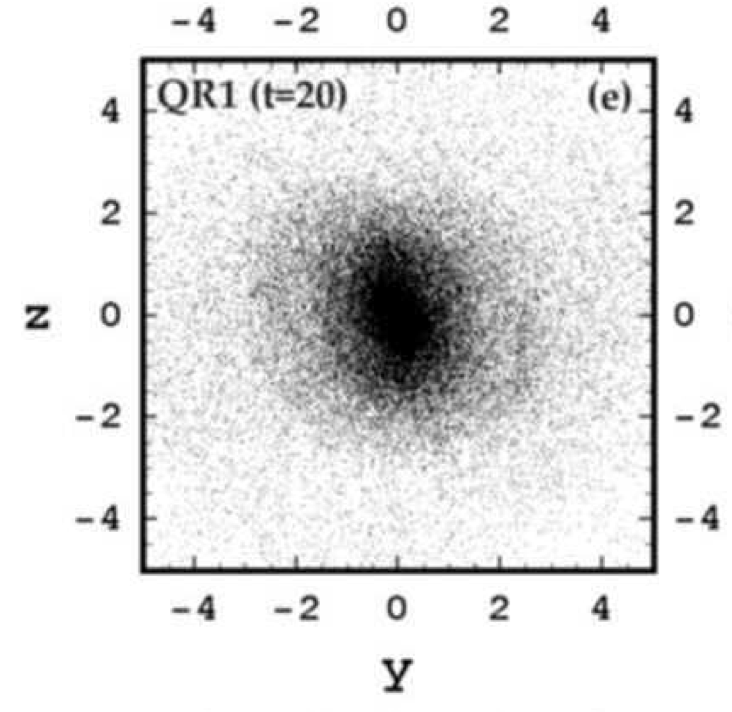}
     \includegraphics[width=0.48\linewidth]{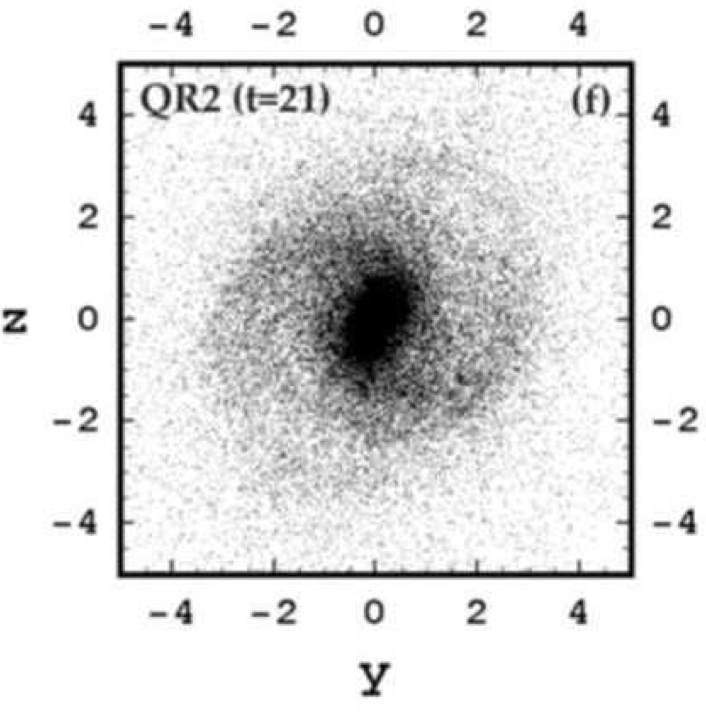}
     \includegraphics[width=0.48\linewidth]{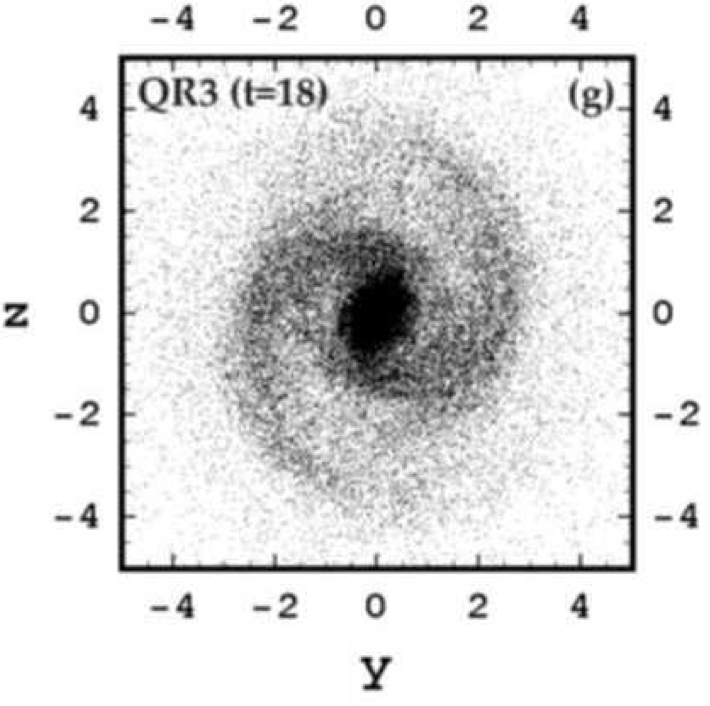}
     \includegraphics[width=0.48\linewidth]{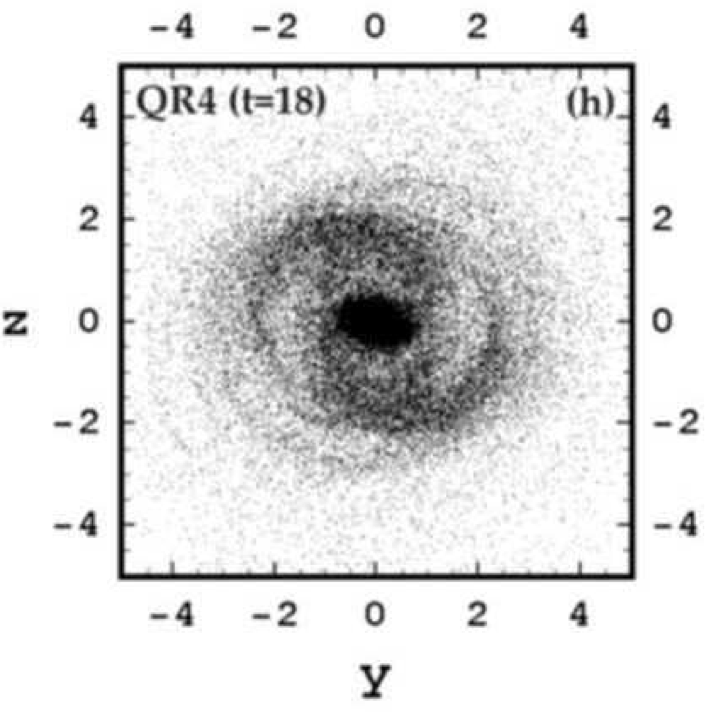}
     \caption{Snapshots   of   the   N-body   simulations  from   Fig.    1   in
       \citet{voglis:2006}. The arm/spiral structures are obtained by increasing
       the total angular momentum  from models QR1 to QR4 for  the same mass and
       size for all the simulations.} \label{fig:qrs}
   \end{centering}
   \end{figure}

\begin{figure}
   \begin{centering}
     \includegraphics[width=1.0\linewidth]{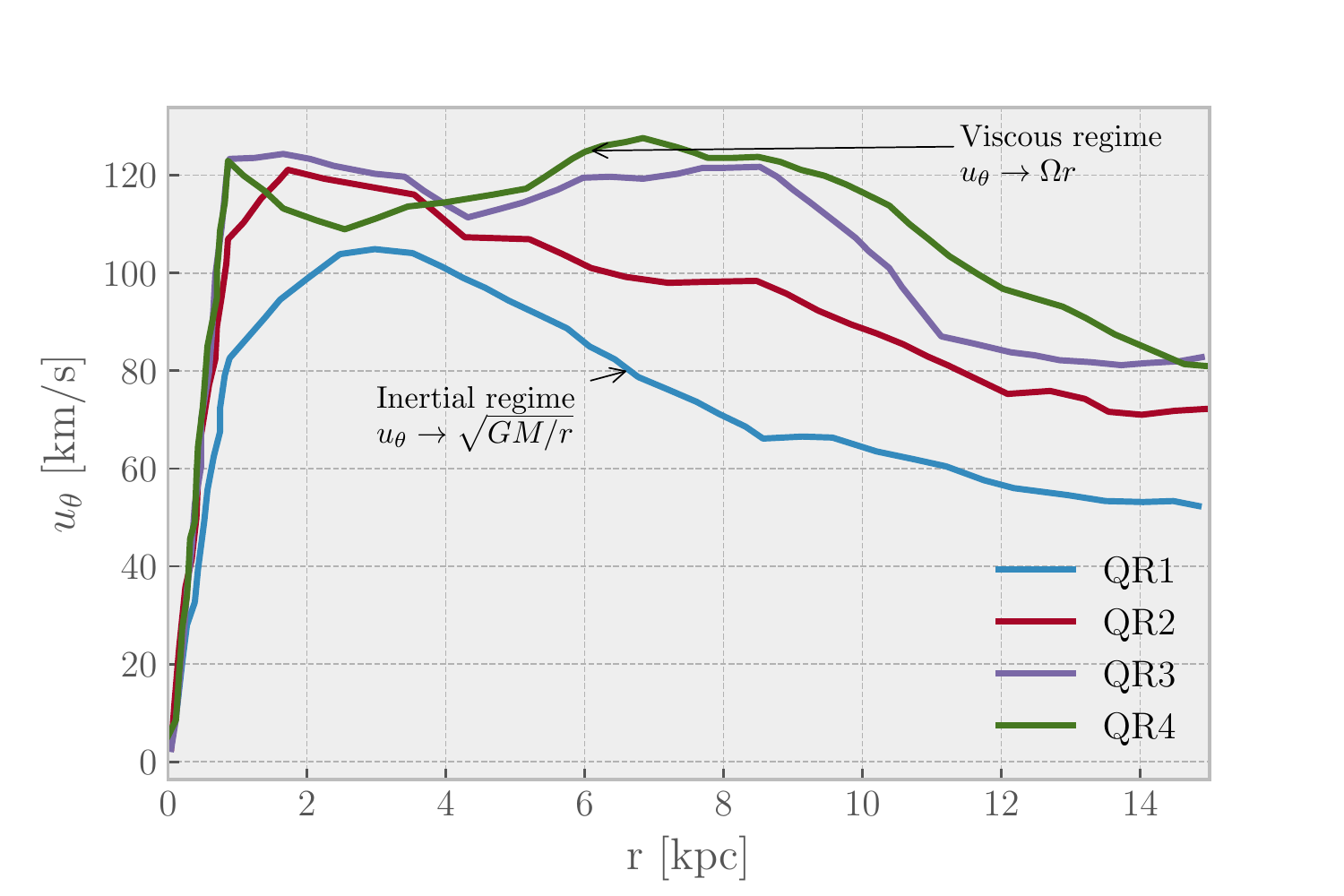}
     \caption{Rotation  curves of  the different  models  from QR1  to QR4  from
       Fig. 2 in \citet{harsoula:2009}.} \label{fig:qrs_rot}
   \end{centering}
   \end{figure}

   In principle, it should be possible  to verify whether this interpretation is
   correct   or  not   by   exploring  such   regimes   with  N-body   numerical
   simulations. The result of such a study is presented in Figure~\ref{fig:qrs},
   from the  N-body simulations of  \citet{voglis:2006} who increased  the total
   angular momentum at fixed  total mass $M$ and size $R$,  which means that the
   gravitational energy $GM/R$ is constant for all the simulations.  One can see
   the emergence of the arm/spiral structures on a scale $\lambda_{arm}$ smaller
   than $R$. If one cannot explore the full N-body dynamics and approximates the
   simulation by  a fluid, the  structure bounded  by the spiral  arms indicates
   that the Jeans  length is smaller than  the size of the  system. This implies
   that a  non-ideal description is  needed when the spiral-arm  structures have
   formed.   In such  a case,  the  gravitational interactions  at small  scales
   inside the spiral  arms has to be  taken into account in  the viscous stress,
   indicating a transition to the viscous regime. Fig.~\ref{fig:qrs_rot}
   displays  the   rotation  curves  associated   with  the  simulations   shown
   in 
   Fig.~\ref{fig:qrs}, assuming a half mass radius of  3 kpc and a total mass of
   10$^{11}$ M$_\odot$ for the galaxy. One can clearly identify a deviation from
   the inertial regime as soon as spiral-arm structures form, which corroborates
   the occurrence of a transition to a viscous regime.

\begin{figure}
   \begin{centering}
     \includegraphics[width=1\linewidth]{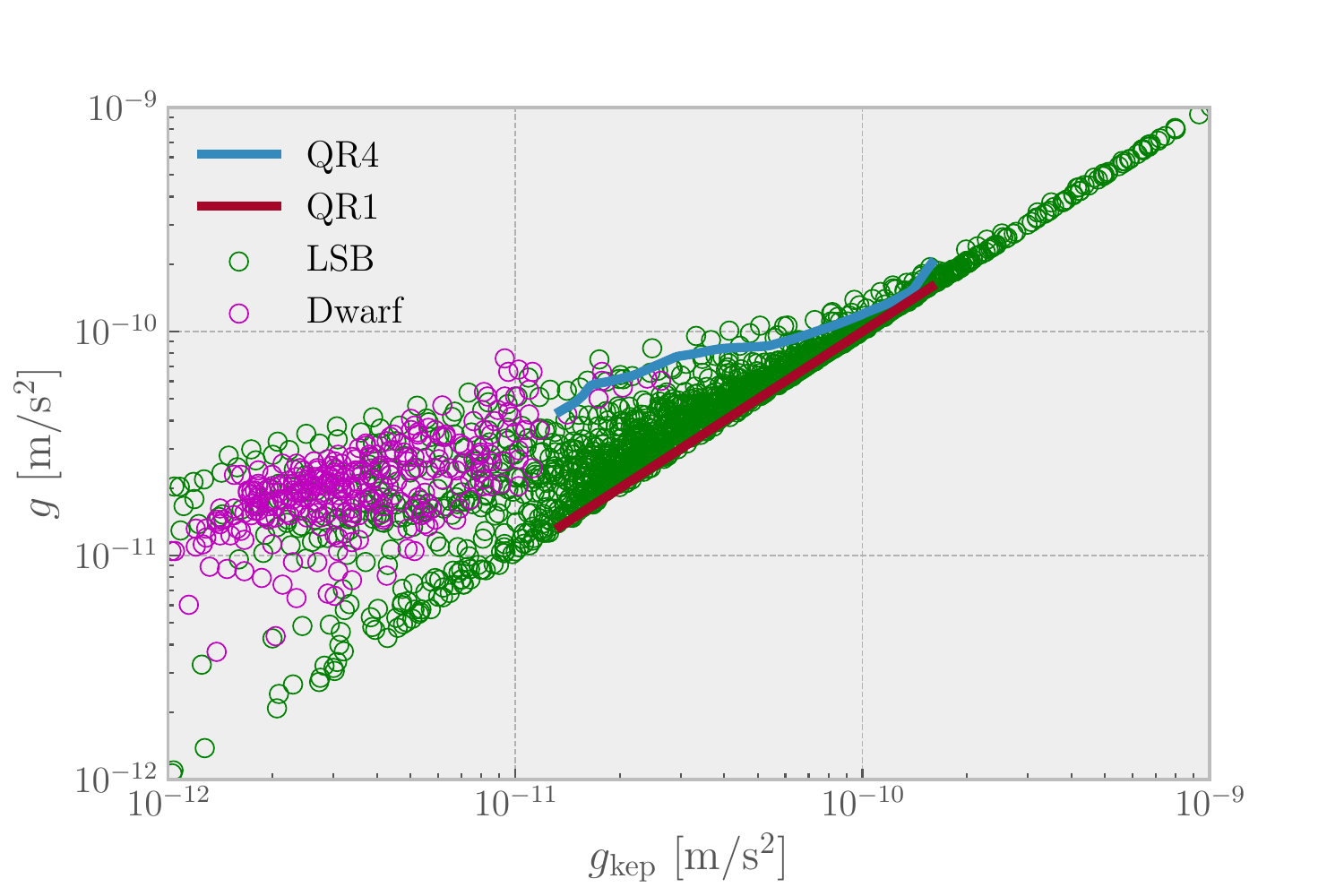}
     \caption{Relation between total acceleration and baryonic component for LSB
       and  dwarf galaxies  \citep{paolo:2019}.  Acceleration  deduced from  the
       rotation  curves of  the QR1  and QR4  models of  \citet{voglis:2006} and
       \citet{harsoula:2009} assuming  a half mass radius  of 3 kpc and  a total
       mass  of  10$^{11}$ M$_\odot$  for  the  galaxy.   QR1  is taken  as  the
       reference  for Keplerian  rotation.   The  bifurcation between  Keplerian
       rotation (QR1) and  ``anomalous'' rotation (QR4) can be  interpreted as a
       phase transition induced  by the presence of  sub-structures i.e.  spiral
       arms that have developed in the QR4 model. }\label{fig:transition}
   \end{centering}
   \end{figure}

   These results  also suggest  that the  change in  the rotation  curve between
   spiral/dwarf  galaxies  and   some  low-surface-brightness  (LSB)  elliptical
   galaxies  exhibiting  Keplerian  profiles  can  be  interpreted  as  a  phase
   transition,  from a  non-viscous to  a viscous  regime, that  depends on  the
   presence  of bound  sub-structures in  the galaxy  (e.g. local  clustering or
   spiral  arms).    An  example  of   such  a  transition  is   illustrated  in
   Fig.~\ref{fig:transition} with the observed data of \citet{paolo:2019}.

   \begin{figure}
   \begin{centering}
     \includegraphics[width=\linewidth]{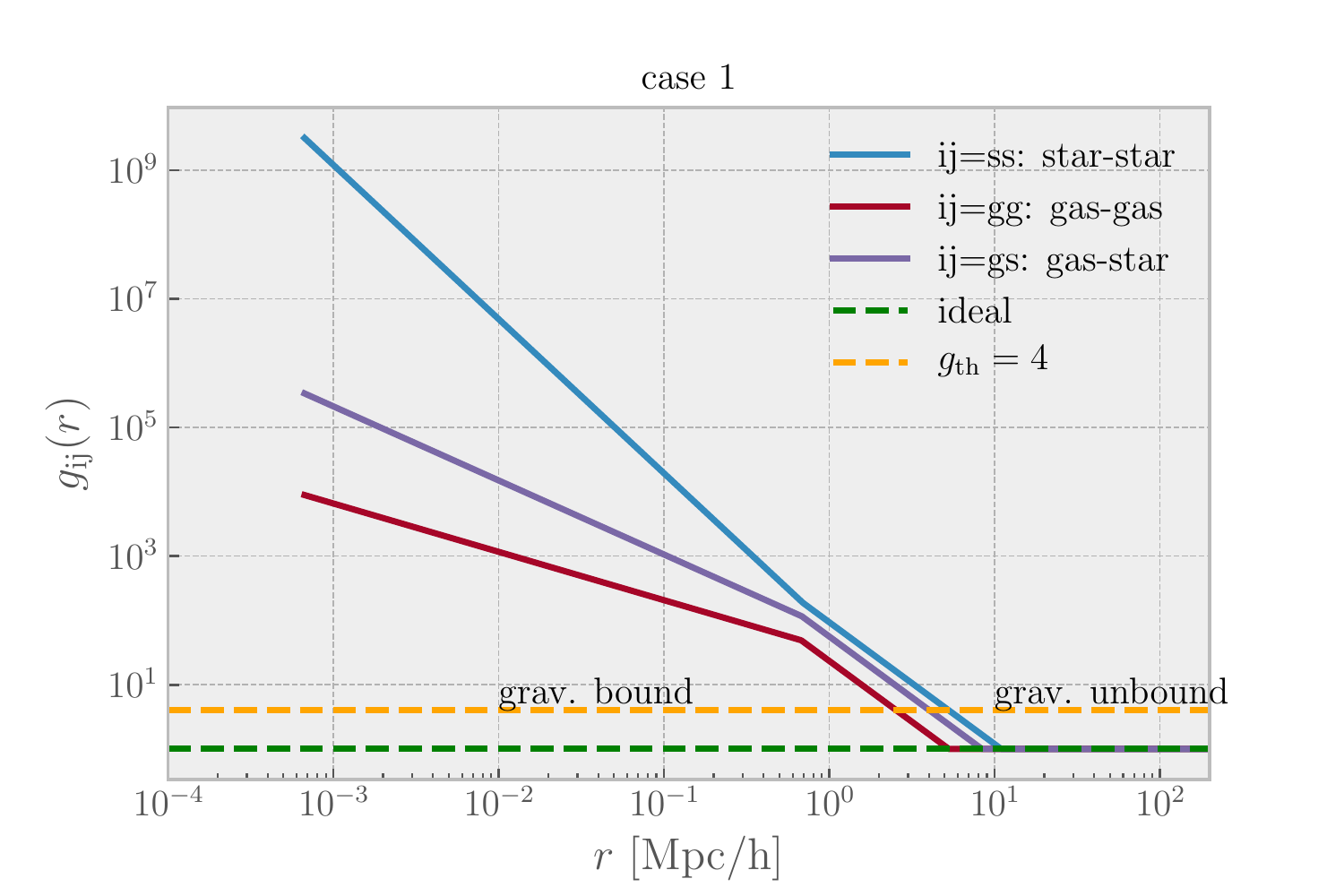}
     \includegraphics[width=\linewidth]{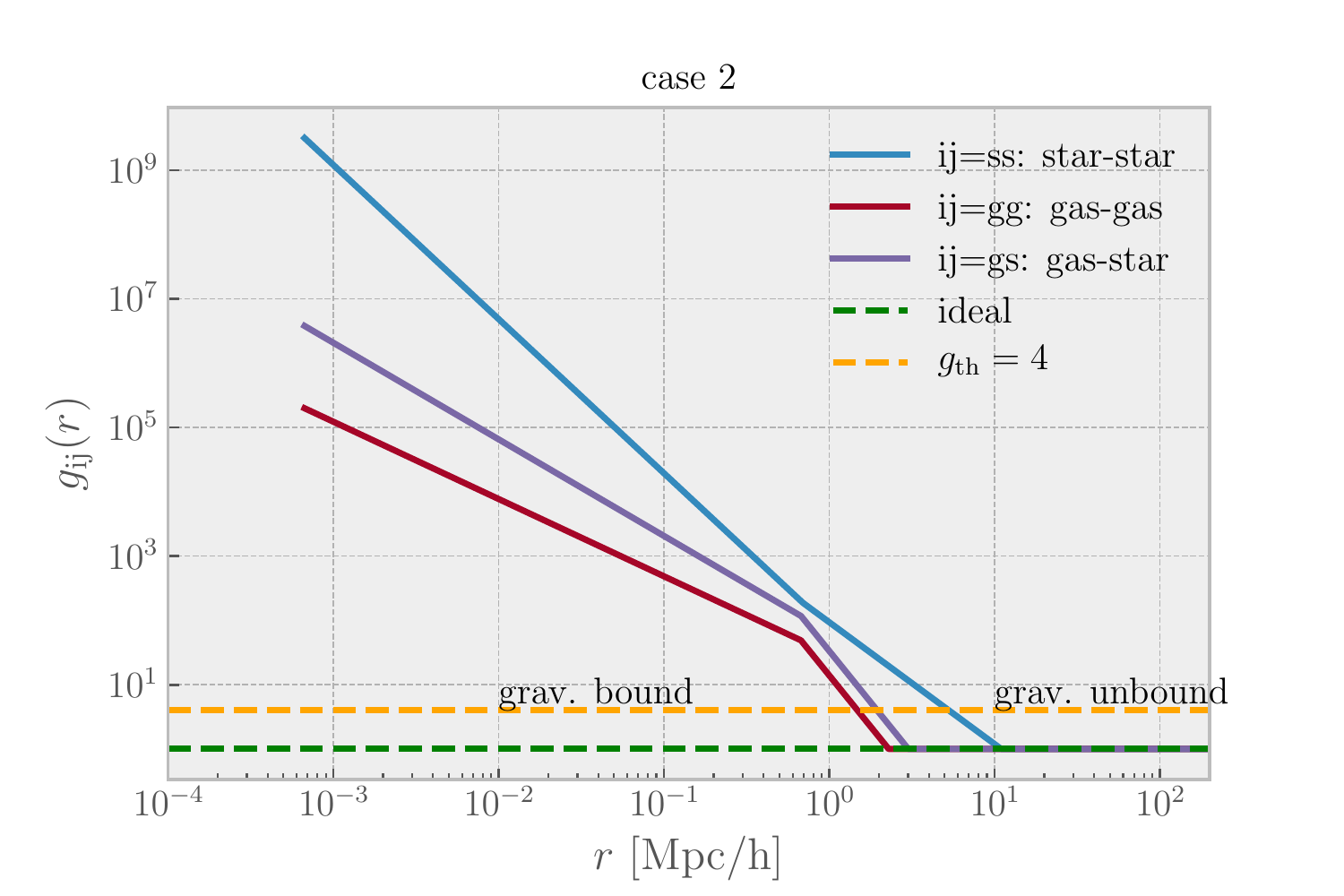}
     \caption{Radial distribution function of stars and gas corresponding to the
       parameters given in Tab.~\ref{tab:gij_param}. Case 1 corresponds to a fit
       of  the  correlation  functions  of the  large-scale  structures  of  the
       Universe   at   redshift   0   in  the   IllustrisTNG   simulation   from
       \citet{springel:2017}. Case  2 uses the  same stellar component  but more
       compact sub-structures in the gas component.} \label{fig:lss_z0}
   \end{centering}
   \end{figure}

   \subsection{Non-ideal cosmology}\label{sec:cosmo}

   As  discussed  in   Sect.~\ref{sec:virial}  and  Sect.~\ref{sec:newton},  the
   Poisson equation  and the equivalence principle  can be safely used  only for
   uncorrelated  fluids. Indeed,  the  Poisson and  Einstein  equations are  not
   designed to account  for (short-scale) correlations that can be  present in a
   fluid. Yet, the  existence of large-scale structures that  have formed during
   the  evolution of  the  Universe means  that  we  are in  the  presence of  a
   correlated  fluid, even  though at  very large  scales, the  Universe can  be
   considered as homogeneous and isotropic.  One  can even argue that, at scales
   characteristic of these structures, the Universe is the most correlated fluid
   in  nature.    Indeed,  as  shown  in   Fig.~\ref{fig:lss_z0}  the  two-point
   correlation  function  and  radial  distribution function  can  reach  values
   $g(r)\approx 10^8$, to  be compare with $g(r)=1$ for  an ideal (uncorrelated)
   fluid  (see Fig.~\ref{fig:lss_z0}).   We show  below how  to incorporate  the
   effect of these correlations into the Friedmann equations. For simplicity, we
   restrict ourselves to the Newtonian limit and we do not intend to explore for
   now an extension  of general relativity to the non-ideal  regime. As such, it
   can be considered  as a preliminary approach  at this stage. As shown in many
   studies and  textbooks \citep[see, e.g.][]{peacock:1999},  however, Newtonian
   cosmology  has proved  to  be  a very  useful  framework  to explore  various
   physical  effects in  cosmology,  while keeping  the calculations  relatively
   simple.

   Assuming a homogeneous  and isotropic universe, we can  obtain the well-known
   Friedmann equations from Einstein's equation

   \begin{eqnarray}
     H^2     +\frac{K}{a^2}      &=&     \frac{8\pi      G}{3c^2}\rho_e,     \cr
     \dot{H}+\frac{3}{2}H^2 + \frac{K}{2 a^2} &=& -\frac{4\pi G}{c^2}P,
   \end{eqnarray}
   with $\rho_e$  and $P$ the  energy density and  pressure, $H=\dot{a}(t)/a(t)$
   the Hubble parameter, $K$ the Gaussian  curvature and $a(t)$ the scale factor
   defining the geometry of the  Universe, with $ds^2=a(t)^2 d\Sigma^2-c^2 dt^2$
   and   $d\Sigma^2=dr^2/(1-Kr^2)+r^2d\Omega^2$.   We    shall   consider   flat
   geometries,   with  $K=0$,   and   we  define   the  deceleration   parameter
   $q=-1-\dot{H}/H^2$.  The Friedmann equations are then given by
   \begin{eqnarray}\label{eq:friedmann}
     H^2 &=& \frac{8\pi G}{3c^2}\rho_b, \cr q &=& -1-\frac{\dot{H}}{H^2} \approx
     \frac{1}{2},
   \end{eqnarray}
   with $\rho_b$ the energy density of  baryons present in the matter era. Based
   on the observed  density of baryons, the expansion from  the Hubble parameter
   should be of the order of 15 km/s/Mpc and the deceleration parameter close to
   1/2. As a  consequence, this model cannot account for  the observed expansion
   of the  Universe close  to 67  km/s/Mpc from  the latest  data of  the Planck
   mission \citep{planck:2018},  and the acceleration of  the expansion measured
   by Type Ia supernovae \citep{riess:1998}, $q=-1.0 \pm 0.4$.

   To  account for  these discrepancies,  the standard  cosmological model  must
   invoke the presence of cold dark matter  (CDM) and dark energy in the form of
   a cosmological constant $\Lambda$, leading to the equations
   \begin{eqnarray}
     H^2   &=&   \frac{8\pi  G}{3c^2}(\rho_b   +\rho_\mathrm{CDM})+\frac{\Lambda
       c^2}{3},\cr       q        &=&       -1-\frac{\dot{H}}{H^2}       \approx
     \frac{1}{2}-\frac{\Lambda c^2}{2H}.
   \end{eqnarray}
   This adjustment leads to the classical energy budget of the Universe in which
   baryonic matter accounts  for $\sim$5\%, cold dark matter  for $\sim$20\% and
   dark energy ($\Lambda$) for $\sim$75\%.

   We will  now show how to  include correlations into the  Friedmann equations.
   As mentioned  above, we  present for  now the  approach within  the Newtonian
   limit to illustrate our point.   Following \citet{peacock:1999}, we assume an
   ensemble of particles of mass $m$ in  a volume $V$ of radius $R$, whose fluid
   density is assumed to be homogeneous  and isotropic at very large scales, but
   can   be  correlated   at  smaller   scales,  i.e.    $P_1(\vec{r})=1/V$  but
   $P_2(\vec{r},\vec{r}^\prime)\ne P_1(\vec{r})P_1(\vec{r}^\prime)$.   The total
   mechanical energy of this system, under its N-body or fluid form, is given by
   (see \S~\ref{sec:virial})
   \begin{eqnarray}\label{eq:emec}
     E_m = \langle\ m v^2/2\rangle + \langle H_\mathrm{int} \rangle.
   \end{eqnarray}
   Similarly  to  $\alpha_{\mathrm{ih}}$   in  Sect.~\ref{sec:appli_virial},  we
   define $\alpha_{\mathrm{ni}}$ by
   \begin{equation}\label{eq:alpha_ni_general}
     \alpha_{\mathrm{ni}}    =     \frac{\iint_{V,V}    \phi(|\vec{r}    -
       \vec{r}^\prime|)     P_2(\vec{r},    \vec{r}^\prime)     dV    dV^\prime}
           {\iint_{V,V}  \phi(|\vec{r}   -  \vec{r}^\prime|)  P_1(\vec{r})
             P_1(\vec{r}^\prime) dV dV^\prime}.
   \end{equation}
   The     ideal    (uncorrelated)     fluid    hypothesis     corresponds    to
   $\alpha_{\mathrm{ni}} = 1$. Eq.~\ref{eq:emec} can then be rewritten as
   \begin{eqnarray}
     E_m   &=&    \langle\   m   v^2/2\rangle    +\alpha_{\mathrm{ni}}   \langle
     H_\mathrm{int,ideal} \rangle.
   \end{eqnarray}
   Following  \citet{peacock:1999},   we  take  $  \langle\   m  v^2/2\rangle  =
   M\dot{R}^2/2$  and  $\langle  H_\mathrm{int,ideal}  \rangle  =  -GM^2/R$  and
   rewrite this equation as
   \begin{eqnarray}
     E_m/M  =  \frac{\dot{R}^2}{2}-  \frac{4\pi}{3}\alpha_{\mathrm{ni}}G  \rho_m
     R^2.
   \end{eqnarray}
   Re-arranging the different terms, we get
   \begin{eqnarray}\label{eq:hubble_ni}
     \frac{\dot{R}^2}{R^2}     -\frac{2E_m/M}{R^2}     =     \frac{8\pi     G}{3
       c^2}\alpha_{\mathrm{ni}} \rho_m c^2
   \end{eqnarray}
   In   the   uncorrelated   case,  $\alpha_{\mathrm{ni}}=1$,   and   with   the
   substitutions $R\rightarrow a$  and $-E_m/M \rightarrow K$,  we recognize the
   first    Friedmann    equation   in    Eq.~\ref{eq:friedmann}.     Therefore,
   Eq.~\ref{eq:hubble_ni} indicates  that correlations  should be  accounted for
   through a multiplicative factor $\alpha_{\mathrm{ni}}$ of the energy density,
   for a  proper non-ideal generalization  of the Friedmann equation  (note that
   this  could also  be seen  as a  multiplicative factor  of the  gravitational
   constant). The last  step in getting a non-ideal first  Friedmann equation is
   to take the thermodynamic limit in $\alpha_{\mathrm{ni}}$, $(N,V) \rightarrow
   \infty$  keeping $N/V=\rho$  constant.  This  can be  done  by introducing  a
   near-field approximation on a scale  $\lambda_{H}$, i.e.  replacing $\phi$ by
   $\exp(-r/\lambda_H)\phi$  in  Eq.~\ref{eq:alpha_ni_general}, or  equivalently
   introducing  a cutoff  of  the integrals  at  a radius  $r_{\mathrm{bound}}$,
   roughly  of the  order of  $\lambda_H$. In  the latter  case, and  assuming a
   large-scale flat,  homogeneous and isotropic  Universe, we get  the following
   first non-ideal Friedmann equation
   \begin{equation}
     H_\mathrm{ni}^2 = \frac{8 \pi G}{3c^2}\alpha_\mathrm{ni} \rho_b
   \end{equation}
   with
   \begin{equation}
     \alpha_\mathrm{ni}     =    \frac{\int_0^{r_\mathrm{bound}}     g(r)r    dr
     }{\int_0^{r_\mathrm{bound}} r dr}.
   \end{equation}

   We can now look at the acceleration  of the expansion.  Since we can link the
   value of the Hubble parameter to the non-ideal amplification induced by bound
   sub-structures, $\alpha_\mathrm{ni}$, it is natural to expect an acceleration
   of the expansion linked to an  increase of the densities of these structures,
   due to  the ongoing  gravitational collapse.  In order  to derive  the second
   non-ideal Friedmann  equation, we use the  first law of thermodynamics  in an
   expanding universe

   \begin{equation}
     \frac{d(\rho_e a^3)}{dt} = -P_\mathrm{ni} \frac{da^3}{dt},
   \end{equation}
   where $P$ is a non-ideal pressure.   This yields the relation $\dot{\rho_e} +
   3 \dot{a}/a \rho_e = - 3P\dot{a}/a $.  Using this relation and the derivation
   of  the first  non-ideal  Friedmann  relation, we  get  the non-ideal  second
   Friedmann equation:
   \begin{equation}
     \dot{H}_\mathrm{ni}  +  \frac{3}{2}H_\mathrm{ni}^2   +  \frac{K}{2  a^2}  =
     \frac{-4\pi   G}{c^2}    P_\mathrm{ni}\alpha_\mathrm{ni}   +   \frac{4\pi    G}{3   c^2
       H_\mathrm{ni}}\rho_e \dot{\alpha_\mathrm{ni}}.
   \end{equation}
   Assuming a flat geometry, $K=0$, we get the deceleration parameter as
   \begin{equation}
     q_\mathrm{ni}       =      \frac{1}{2}       +      \frac{4\pi       G}{c^2
       H_\mathrm{ni}}P_\mathrm{ni}\alpha_\mathrm{ni}                                       -
     \frac{\dot{\alpha_\mathrm{ni}}}{2H_\mathrm{ni} \alpha_\mathrm{ni}}.
   \end{equation}
   The non-ideal pressure term is negligible  as in the ideal case.  However, we
   can get  a negative  deceleration parameter  with the  third term  because of
   $\dot{\alpha_\mathrm{ni}}$,  which   is  positive  since  the   densities  of
   sub-structures  are increasing  with time  due to  the ongoing  gravitational
   collapse.  This  term can then  replace the contribution of  the cosmological
   constant that  needs to be introduced  in the Friedmann equations  to explain
   the acceleration of the expansion in the standard $\Lambda$CDM approach.

   To  summarize, the  two non-ideal  Friedmann equations  (for $K=0$) in our  formalism are
   given by:
   \begin{eqnarray}
     H_\mathrm{ni}^2  &=&  \frac{8   \pi  G}{3c^2}\alpha_\mathrm{ni}  \rho_b,\cr
     q_\mathrm{ni}  &=&  -1-\frac{\dot{H}_\mathrm{ni}}{H_\mathrm{ni}^2}  \approx
     \frac{1}{2}         -        \frac{\dot{\alpha_\mathrm{ni}}}{2H_\mathrm{ni}
       \alpha_\mathrm{ni}},
   \end{eqnarray}
   with
   \begin{equation}
     \alpha_\mathrm{ni}     =    \frac{\int_0^{r_\mathrm{bound}}     g(r)r    dr
     }{\int_0^{r_\mathrm{bound}} r dr}.
   \end{equation}
   We can  now estimate  $g(r)$ from available  numerical simulations.  We first
   decompose  $\alpha_\mathrm{ni}$  into  the  dominant  components  of  visible
   matter, namely gas and stars:
   \begin{equation}
     \alpha_\mathrm{ni}     =     X_\mathrm{star}^2     \alpha_\mathrm{ss}     +
     (1-X_\mathrm{star})^2           \alpha_\mathrm{gg}            +           2
     X_\mathrm{star}(1-X_\mathrm{star}) \alpha_\mathrm{gs},
   \end{equation}
   where $ss,  gg, sg$  denote the star-star,  gas-gas and  star-gas interaction
   contributions, respectively, and $X_\mathrm{star}$  denotes the mass fraction
   in  stars relative  to the  total visible  baryonic matter  in gas  and stars
   ($X_\mathrm{star} \approx 5 \%$). We can parameterize the radial distribution
   functions for gas and stars from the auto- and cross-correlation functions of
   the  large-scale  structures of  the Universe  \citep{peebles:1980} by  using
   simulations  at  redshift  $z=0$ from  \citep{springel:2017}\footnote{As  the
     structures obtained in $\Lambda$CDM cosmological simulations represent well
     the observations,  we use them  as the  reference in our  calculations.  In
     contrast to the  conventional $\Lambda$CDM model, however,  dark matter and
     dark energy in  our formalism are in reality proxies  for non-ideal effects
     induced by sub-structures.}, as illustrated in Fig.~\ref{fig:lss_z0}. It is
   important to stress that the  decomposition $\alpha_\mathrm{ni}$ must be made
   with {\it squared} mass fractions  and account for cross-correlations because
   the  interaction energy  depends  formally on  $P_2(\vec{r},\vec{r}^\prime)$,
   that can  be seen  as the  square of the  density \citep[similarly  to Eq.~15
     in][]{springel:2017}.  We recall the  link between the correlation function
   and  the   radial  distribution   function  $\xi(r)  =g(r)-1$.    The  radial
   distribution functions can be parametrized with the simple forms
   \begin{eqnarray}
     g_\mathrm{ij}(r)       &=      \max       \left(g_\mathrm{1Mpc}\times\left(
     \frac{r}{\mathrm{1Mpc}}\right)^{-\beta_0},1\right),                   \quad
     r<\mathrm{1Mpc}\cr         &=         \max\left(g_\mathrm{1Mpc}\times\left(
     \frac{r}{\mathrm{1Mpc}}\right)^{-\beta_1},1\right), \quad r\ge\mathrm{1Mpc}
   \end{eqnarray}
   We assume for  simplification that the radial distribution  function is equal
   to  1 at  large  distances, while  strictly speaking  it  should be  slightly
   smaller than 1 and tend to  1 at infinity.  The parameters $g_\mathrm{1Mpc}$,
   $\beta_0$  and $\beta_1$  are  given in  Table  ~\ref{tab:gij_param} for  two
   cases.  Case 1 corresponds to the  fit of the correlation functions presented
   in  \citet{springel:2017},  assuming  $\xi(r) \approx  g(r)$  for  $\xi(r)\gg
   1$. Case 2 has  the same parametrization of the stellar  component as case 1,
   is  well constrained  by observations  \citep[e.g.][]{li:2009}, but  has more
   compact sub-structures in the gas  component. This modification is ad-hoc for
   the moment  and further investigations are  needed to check if  its amplitude
   would be compatible with the  observational constrains that are available for
   the gas  component.  The  main point of  case 2, however,  is to  explore the
   sensitivity   of    the   non-ideal   amplification   to    the   {\it   gas}
   sub-structures. \\ We then define  the bound radius $r_\mathrm{bound}$ as the
   minimal    radius   such    that   $g_\mathrm{ij}(r)<    g_\mathrm{th}$   for
   $r>r_\mathrm{bound}$  and we  explore  two values  for  the threshold  value:
   $g_\mathrm{th}=1$ and $g_\mathrm{th}=4$.

   \begin{table}
   \centering
   \begin{tabular}{l|l|l|l}
     &  \multicolumn{3}{c}{case  1}  \\  \hline  \hline  &  $g_\mathrm{1Mpc}$  &
     $\beta_0$  &  $\beta_1$  \\  \hline  $g_\mathrm{gg}$ &  50  &  0.75  &  1.9
     \\ $g_\mathrm{gs}$ & 120 & 1.15 & 1.9 \\ $g_\mathrm{ss}$ & 200 & 2.40 & 1.9
     \\
   \end{tabular}
   \begin{tabular}{l|l|l|l}
     &  \multicolumn{3}{c}{case  2}  \\  \hline  \hline  &  $g_\mathrm{1Mpc}$  &
     $\beta_0$  &  $\beta_1$  \\  \hline  $g_\mathrm{gg}$  &  50  &  1.2  &  3.2
     \\ $g_\mathrm{gs}$ & 120  & 1.5 & 3.2 \\ $g_\mathrm{ss}$ & 200  & 2.4 & 1.9
     \\
   \end{tabular}
   \caption{ Parameters used for the  parametrization of the radial distribution
     function of gas and  stars for case 1 and case  2. The corresponding radial
     distribution           functions          are           plotted          in
     Fig.~\ref{fig:lss_z0}.}\label{tab:gij_param}
   \end{table}

   \begin{table}
   \centering
   \begin{tabular}{l|l|l|l|l}
     &   \multicolumn{4}{c}{case  1}   \\   \hline   \hline  $g_\mathrm{th}$   &
     \multicolumn{2}{c|}{1}      &      \multicolumn{2}{c}{4}     \\      \hline
     $\alpha_\mathrm{gg}$         &          \multicolumn{2}{c|}{5.02}         &
     \multicolumn{2}{c}{15.56}         \\         $\alpha_\mathrm{gs}$         &
     \multicolumn{2}{c|}{6.28}             &            \multicolumn{2}{c}{21.0}
     \\       $\alpha_\mathrm{ss}$      &       \multicolumn{2}{c|}{61.0}      &
     \multicolumn{2}{c}{257} \\ \hline  \hline $X_\mathrm{star}$ & 0.05  & 0.1 &
     0.05 &  0.1 \\  \hline $\alpha_\mathrm{ni}$  & 5.28  & 5.80  & 16.7  & 18.9
     \\   $H_\mathrm{ni}$  &   34.2   &   35.8  &   60.7   &   64.7  \\   \hline
     $\dot{\alpha}_\mathrm{ni}$ & 0.12 & 0.13 & 0.20 & 0.21 \\ $q_\mathrm{ni}$ &
     -1.20 & -1.21 & -1.12 & -1.09 \\
   \end{tabular}
   \begin{tabular}{l|l|l|l|l}
     &   \multicolumn{4}{c}{case  2}   \\   \hline   \hline  $g_\mathrm{th}$   &
     \multicolumn{2}{c|}{1}      &      \multicolumn{2}{c}{4}     \\      \hline
     $\alpha_\mathrm{gg}$ & \multicolumn{2}{c|}{16.4} & \multicolumn{2}{c}{36.2}
     \\       $\alpha_\mathrm{gs}$      &       \multicolumn{2}{c|}{31.7}      &
     \multicolumn{2}{c}{72.7}         \\          $\alpha_\mathrm{ss}$         &
     \multicolumn{2}{c|}{61.0}  &   \multicolumn{2}{c}{257}  \\   \hline  \hline
     $X_\mathrm{star}$ & 0.05 & 0.1 &  0.05 & 0.1 \\ \hline $\alpha_\mathrm{ni}$
     & 17.9 & 19.6 & 40.2 & 45.0 \\  $H_\mathrm{ni}$ & 62.9 & 65.8 & 94.3 & 94.3
     \\  \hline  $\dot{\alpha}_\mathrm{ni}$  &  0.21   &  0.21  &  0.26  &  0.27
     \\ $q_\mathrm{ni}$ & -1.09 & -1.09 & -0.87 & -0.83 \\
   \end{tabular}
   \caption{Non-ideal  amplification   of  the  different   component  (gas-gas,
     star-star,  and  gas-star)  for  two  different  values  of  the  threshold
     $g_\mathrm{th}$  used   to  define   the  bound  radius.   Total  non-ideal
     amplification $\alpha_\mathrm{ni}$  and the corresponding  non-ideal Hubble
     parameter ($H_\mathrm{ni}$ in  units of km/s/Mpc) for  two different values
     of the star mass fraction relative to the total baryon mass in the Universe
     (we  assume   a  present-day   baryon  density  of   0.25  particle/m$^3$).
     Estimation of $\dot{\alpha}_\mathrm{ni}$ assuming a 14-Gyr-old Universe and
     the       corresponding       non-ideal       deceleration       parameter
     $q_\mathrm{ni}$.}\label{tab:hq_ni}
   \end{table}

   Table \ref{tab:hq_ni} gives the  obtained non-ideal amplification factors for
   the different components  for the two threshold values, as  well as the total
   non-ideal  value  $\alpha_\mathrm{ni}$.   We   also  give  the  corresponding
   non-ideal Hubble parameter for  star mass fractions $X_\mathrm{star}=5\%$ and
   $X_\mathrm{star}=10\%$, respectively.  In the  most conservative case (case 1
   with $g_\mathrm{th}=1$), we get an overall amplification factor between 5 and
   6.    With  a   modification  of   the  gas   sub-structures  (case   2  with
   $g_\mathrm{th}=1$) or a different  threshold (case 1 with $g_\mathrm{th}=4$),
   we get a total amplification of about  20, which yields the observed value of
   the  Hubble parameter  \citep[about 67  km/s/Mpc, ][]{planck:2018}.  The last
   case is extreme (case 2 with  $g_\mathrm{th}=4$) and shows that the non-ideal
   model  can give  a very  rapid expansion,  depending on  the contribution  of
   substructures to the gravitational energy.

   It is also interesting to look  at the non-ideal amplification per component:
   since the stars  are a lot more substructured (correlated)  than the gas, the
   non-ideal amplification for  stars is between 5 and 20  times larger than the
   one for the gas. This effect  could thus explain observations like the bullet
   cluster \citep{markevitch:2004}: when stars  and gas are spatially separated,
   the non-ideal amplification in the  stellar component is significantly larger
   than  the one  in the  gas.  If  the  difference is  around 20  or more,  the
   non-ideal  amplification can  compensate  the difference  in mass  fractions,
   potentially explaining  the observed high gravitational  weak-lensing related
   to the stellar component.

   We    also     give    in     Table    \ref{tab:hq_ni}     estimations    for
   $\dot{\alpha}_\mathrm{ni}$  and  the   corresponding  deceleration  parameter
   $q_\mathrm{ni}$, assuming $\dot{\alpha}_\mathrm{ni}/\alpha_\mathrm{ni}\approx
   \ln(\alpha_\mathrm{ni})/t_u$,  with  $t_u \approx  14$  Gyr  the age  of  the
   Universe.  We see  that the  deceleration parameter  is always  close to  the
   observed value $q = -1.0  \pm 0.4$ \citep{riess:1998}. An interesting feature
   of this  non-ideal model is  that, contrary  to a $\Lambda$CDM  approach that
   needs  to  introduce  two  parameters $\rho_\mathrm{CDM}$  and  $\Lambda$  to
   explain the expansion  and its acceleration, we only  introduce one parameter
   $\alpha_\mathrm{ni}$. We can therefore link the expansion to its acceleration
   and provide an analytical expression  for $q_\mathrm{ni}$, using the critical
   density $\rho_c=3H^2/8\pi  G$.  Assuming  the non-ideal  amplification factor
   $\alpha_\mathrm{ni}$  explains entirely  the observed  present-day expansion,
   i.e.   has  varied  from   $\alpha_\mathrm{ni}=1$  to  $\alpha_\mathrm{ni}  =
   \rho_c/\rho_b \approx 20$ over the age  of the universe, $t_u \sim$14 Gyr, we
   can get an order-of-magnitude analytical expression:
   \begin{equation}\label{eq:qni_rough}
     q_\mathrm{ni} \approx \frac{1}{2} - \frac{\ln(\rho_c/\rho_b)}{2 t_u H},
   \end{equation}
   which gives a deceleration  parameter $q_\mathrm{ni} =-1.06$, compatible with
   the value based on type Ia supernovae \citep[$q=-1.0\pm0.4$, ][]{riess:1998}.

   This  value is  in  tension  with current  estimates  using all  cosmological
   constrains \citep[$q=-0.52\pm0.07$,  ][]{planck:2018}.  It must  be stressed,
   however,  that the  aforementioned  estimates of  the deceleration  parameter
   based on the Planck results rely on the $\Lambda$CDM model. Using the present
   non-ideal cosmological  formalism might  relax this  tension: e.g.   an older
   universe with  an age of  about 20 Gyr  would give a  non-ideal deceleration
   parameter  around  -0.5.  Another  possibility  is  that the  formation  of
   structures has  slowed down in the  recent universe, because e.g  of feedback
   processes. Indeed, observations show that  the cosmic star formation rate has
   decreased by almost  a factor 10 since redshift $z\sim$1-2,  i.e.  within the
   past  $\sim  8$-10  Gyr  \citep{madau:2014}. An  increase  of  the  non-ideal
   amplification factor of about 2 within the  last 7 Gyr would put it in
   agreement  with  current estimates  of  $q$.   Such  a possibility  could  be
   explored in  detail with numerical  simulations. Nonetheless, given  the fact
   that the physics responsible for the  acceleration of the Universe remains so
   far  mysterious, it  is quite  remarkable that  we can  get a  good order  of
   magnitude estimate of the acceleration of the expansion based on a physically
   intuitive and  appealing argument, namely  the increasing density  induced by
   the gravitational collapse of the large-scale structures.

   \section{Discussion and conclusion} \label{sec:conclusions}

   \subsection{A non-ideal Einstein equation?}

   As  discussed in  Sect.~\ref{sec:virial} and  Sect.~\ref{sec:newton}, we  may
   have to  step away  from the  traditional forms of  the Poisson  and Einstein
   equations to get  the non-ideal Friedmann equations.   For a non-relativistic
   fluid, the  mass energy density,  $E_m=\rho c^2$,  is dominant in  the zeroth
   component of the stress energy tensor, $T_{00}\approx \rho c^2$, and it would
   also dominate  all other  forms of  energy, kinetic  and potential:  $E_m \gg
   E_\mathrm{kin}$, $E_p$,  etc... At  first sight, it  seems impossible  for an
   interaction potential energy $E_p$ to dominate over the mass energy, implying
   that correlations  in the interaction  energy should not modify  the Einstein
   equations. It must  be stressed, however, that in  the non-relativistic limit
   of the Einstein equations, the gravitational interaction energy has a special
   status and should not be identified  with $E_p$, but rather directly with the
   integral of $T_{00}$. This can be  seen by deriving the Poisson equation from
   the  Einstein  equations: $\Delta  \phi  \approx  -4\pi G  T_{00}/c^2$.   The
   gravitational interaction energy in the  non-relativistic limit is thus given
   by

   \begin{eqnarray}
     H_\mathrm{int}  &=&  \frac{1}{2}\int_V  \rho(\vec{r})  \phi(\vec{r})  dV\cr
     &\approx&                  -                 \frac{1}{2}\iint_{V,V} G
     \frac{\rho(\vec{r})T_{00}(\vec{r}^\prime)/c^2
     }{||\vec{r}-\vec{r}^\prime||}dVdV^\prime,
   \end{eqnarray}
   which  corresponds to  the uncorrelated  (ideal) version  of the  interaction
   energy. Indeed,  it ignores the  distance between  pairs of particles  in the
   fluid since  it does not  depend on the correlation  function. Heuristically,
   this explains  why the contribution  to the gravitational  interaction energy
   arising from correlations should be  introduced as a multiplicative factor of
   the mass energy  (or of the gravitational constant) and  not as an additional
   term.

   An  other  way  to  see  the  limits  of  using  the  Einstein  equation  for
   self-gravitating correlated fluids is to  look at the relativistic version of
   the      non-ideal      Navier-Stokes       equations      introduced      in
   Sect.~\ref{sec:navier-stockes}:  $\nabla_\mu T^{\mu\nu}=0$.  As explained  in
   Sect.~\ref{sec:navier-stockes}, as soon  as the Jeans length  is not resolved
   (hence when  the cosmological principle is  applied at large scale),  part of
   the gravitational interactions should be taken into account in the EOS of the
   fluid, hence in the definition of $T^{\mu\nu}$. The rest can be accounted for
   as  external interactions  with a  mean  field that  can be  included in  the
   covariant derivative $\nabla_\mu$  in the context of  General Relativity. The
   gravitational interactions  that contribute to  the EOS cannot in  general be
   included in  the covariant derivative, which  is another way to  see that the
   equivalence principle for all the gravitational interactions is broken by the
   presence of correlations in a self-gravitating non-ideal fluid.

   We stress that the concept of non-ideal self-gravity developed in the present
   paper  does not  modify  the fundamental  law of  gravity  between two  point
   particles,  which relies  only on  Newton's  law.  Furthermore,  it does  not
   contradict Poisson and Einstein theories: they always remain valid if one can
   have  access   to  or   compute  exactly   the  full   inhomogeneous  density
   distributions down  to the  scale below  which we  can assume  the gas  to be
   uncorrelated     (see    the     example    of     polytropic    stars     in
   \S\ref{sec:appli_virial}).  However,  for a large ensemble  of particles with
   unresolved  small-scale  inhomogeneities,  a non-ideal  modification  of  the
   Poisson and  Einstein equations is  required to account for  the correlations
   between these sub-structures. It is worth pointing out that a modification of
   the Poisson equation  because of correlations is a procedure  already used in
   the case  of the  electrostatic force  in the  context of  physical chemistry
   \citep[see,  e.g.][]{storey:2012}.  In  this   field,  Poisson  and  Einstein
   theories would be  qualified as mean field theories, i.e.   only valid in the
   uncorrelated case.
   A more complete theoretical approach to develop a non-ideal formalism for the
   Einstein equations could come from  two routes.  Either from a homogenization
   procedure of the  inhomogeneous Einstein equations, a  path already currently
   explored \citep[see, e.g.][]{buchert:2000,buchert:2008}, or from an extension
   of the  concepts used in statistical  mechanics, as developed in  this paper,
   within a general relativistic framework.

   \subsection{Conclusion}

   Starting  from the  well-established  BBGKY formalism,  we  have shown  that
   correlations induced  by bound  sub-structures should  be accounted  for when
   describing the energy-mass density of the Universe. This  should be done
   via  the  use   of  a  non-ideal  Virial  theorem   formalism  and  non-ideal
   Navier-Stokes equations.  Taking into account these non-ideal effects yields
   an  "amplification"  of  the  gravitational interaction  energy  which  could
   account, at least partly, for the  missing mass problem in galaxies, clusters
   of  galaxies, and  large-scale structures  in general.  The strength  of this
   model is that the radial  distribution function (or the two-point correlation
   function) can be  well determined in galaxies and  the large-scale structures
   either by observations or by simulations.

   By looking  at the  viscous limit of  the non-ideal  Navier-Stokes equations
   induced  by  the presence  of  small-scale  interactions,  we show  that  the
   presence  of sub-structures  (e.g.   spiral arms  and  local clustering)  can
   produce non-Keplerian rotation profiles that  could explain the observed flat
   rotation curves. A consequence of this  approach is to show that the observed
   bifurcation  between  spiral/dwarf  galaxies   and  some  LSB  galaxies  with
   Keplerian rotation profiles, could be explained by the presence or absence of
   sub-structures within these galaxies. In the context of statistical mechanics
   this  bifurcation could  be interpreted  as  a phase  transition between  two
   different equilibrium (viscous and non-viscous) states.

   Using  this non-ideal  Virial theorem,  we have  derived non-ideal  Friedmann
   equations,  within  the  Newtonian  limit.   Using  the  radial  distribution
   function of visible gas and stars obtained in large-scale simulations, we can
   compute the non-ideal  amplification factor of gravitational  mass energy and
   show that this  factor can easily be of  the order of 5 to 20  and thus could
   explain the observed value of the Hubble parameter. Furthermore, we show that
   the amplification is much stronger for the stellar component than for the gas
   component, which  could also  explain the bullet  cluster observation  with a
   high gravitational weak lensing in the  stellar component compared to the gas
   component.

   Furthermore,  since most  of  the contribution  to the  value  of the  Hubble
   parameter stems from  the non-ideal amplification caused  by the interactions
   between  sub-structures,  this naturally  explains  the  acceleration of  the
   expansion of the universe, as a consequence of the ongoing collapse, thus the
   increasing  densities  of  these  sub-structures.   An  amplification  factor
   $\alpha_{\rm ni}\simeq 20$ during most of the lifetime of the Universe yields
   a deceleration parameter $q_{\rm  ni}\simeq -1$ (see Eq. \ref{eq:qni_rough}),
   close  to the  estimates based  on type  Ia supernovae.  To the  best of  our
   knowledge, this model is  the first one able to predict  a coherent value for
   the  acceleration  of the  expansion  of  the universe  with  first-principle
   physical arguments.

   Observations  of the  {\it Euclid}  satellite will  be crucial  to probe  and
   further constrain the radial distribution  function and the non-ideal effects
   of self-gravitating  matter at  large scale.   A precise  characterization of
   these effects will be essential to  build a full non-ideal cosmological model
   addressing  all  cosmological  constrains  (e.g.  CMB  data)  and  to  reveal
   eventually what is really dark in our Universe.

   \begin{acknowledgements}
     We thank the anonymous referee  for his/her constructive comments. We thank
     S.  Kokh and E.  Audit for co-supervising the PhD of T. Padioleau. This PhD
     subject  is  at  the  frontier  between  astrophysics  and  two-phase  flow
     simulations for  the cooling system  of nuclear  power plants and  the link
     between liquid statistical mechanics and gravity at large scale has emerged
     from this trans-disciplinary approach.  They also thank CEA in general (DRF
     and DES  in particular)  for the  funding of  the PhD  and for  providing a
     favorable environment for trans-disciplinarity  at Maison de la Simulation.
     PT also thanks  the Astrosim 2019 school and in  particular J.  Rosdhal and
     O.  Hahn for their lectures: part of  the ideas presented in this paper has
     emerged after  the school.  The  authors also thank  F. Sainsbury-Martinez,
     J.  Faure,  T.  Buchert,  G.   Laibe,  Q.   Vigneron,  P.   Hennebelle,  F.
     Bournaud,  D.  Elbaz,  D.  Borgis,  R.   Balian, and  L. Saint-Raymond  for
     interesting discussions and useful comments on this paper.  PT acknowledges
     supports  by  the European  Research  Council  under Grant  Agreement  ATMO
     757858.
   \end{acknowledgements}

\bibliographystyle{aa} \bibliography{main}

\end{document}